\documentclass[pdflatex,sn-mathphys]{sn-jnl}

\jyear{2022}%

\theoremstyle{thmstyleone}%
\theoremstyle{thmstyletwo}%

\theoremstyle{thmstylethree}%

\begin{document}

\title{Quantum State Transfer: Interplay between Gate and Readout Errors}

\author*[1]{\fnm{Bharat} \sur{Thotakura}}\email{bharat.thotakura@stonybrook.edu}

\author[2]{\fnm{Tzu-Chieh} \sur{Wei}}

\affil[1,2]{\orgdiv{Department of Physics and Astronomy}, \orgname{State University of New York at Stony Brook}, \orgaddress{\postcode{ 11794-3840}, \state{NY}}}

\affil[2]{\orgdiv{C. N. Yang Institute for Theoretical Physics}, \orgname{State University of New York at Stony Brook}, \orgaddress{\postcode{ 11794-3840}, \state{NY}}}


\abstract{Quantum networks consist of quantum nodes that are linked by entanglement and quantum  information can be transferred from one node to another. Operations can be applied to qubits of local nodes coordinated by classical communication to manipulate  quantum states  and  readout/measurement will be employed to obtain results. Here, we use quantum circuits to simulate quantum state transfer between two nodes connected in a linear geometry through other nodes. We explore the interplay between gate and readout errors on the performance of state transfer. We find that the nominal success probability is not necessarily a monotonic function of the two error rates and employ numerical simulations and analytic tools to understand their interplay.}


\maketitle
\newpage
\section{Introduction} \label{sec:intro}
One of the goals in quantum technology research deals with linking up many local quantum devices to act together as a network, called a quantum network or internet. The local nodes in that network  (see Ref.~\cite{kimble2008quantum}) are quantum processing units, which can be thought of as small quantum computers with quantum memory. The links connecting these nodes can be either classical or quantum channels (or both)---that is, equipped with a mechanism of communication between nodes, with the basic protocols to establish entanglement, such as quantum teleportation~\cite{bennett1993teleporting} and entanglement swapping \cite{pan1998experimental}. 

In 2001, Duan, Lukin, Cirac, and Zoller (DLCZ) proposed a long-distance quantum communication protocol (i.e., the DLCZ protocol) over long lossy channels  using photons and atomic ensembles~\cite{duan2001long}. Despite the simplicity of the scheme, compared to the complexity of realizing a universal quantum computer, a full scale implementation of the DLCZ protocol over long distances has not yet been carried out. However,  advancements are being made,  such as individual addressability of multiplexed quantum memory ~\cite{pu2017experimental}, teleportation between non-neighboring nodes~\cite{hermans2022qubit}, the distribution of Greenberger–Horne–Zeilinger (GHZ) state and multi-node entanglement swapping (see Ref.~\cite{pompili2021realization}), and establishment of long-range entanglement between single atoms~\cite{van2022entangling}. Other efforts include  schemes and experiments for remote entanglement generation and distribution towards the development of  stacks of quantum networks~\cite{dahlberg2019link, van2017multiplexed, kozlowski2020designing, humphreys2018deterministic, rozpkedek2019near}. 

As Wehner, Elkouss, and Hanson  lay out in  Ref.~\cite{wehner2018quantum}, all these quantum network stages and efforts coalesce into a unified framework of stages towards a quantum internet. Ultimately, we would like to reach a full quantum computation stage in the quantum network -- before which we would ideally have a few-qubit fault-tolerant network. However, the current stage in development of  quantum computers is more accurately characterized as the NISQ (noisy intermediate-scale quantum) era~\cite{Preskill_2018}, and presently, there are mostly proposals and small-scale tests of fault-tolerant quantum computation~\cite{stephens2013hybrid, campbell2017roads, bourassa2021blueprint, linke2017fault, chen2021exponential} only. Thus, the issue of noise remains one of the biggest impediments to practical and scalable quantum computation, and consequently to the quality and performance of a quantum network. 

In the context of the quantum internet roadmap~\cite{wehner2018quantum}, noise and errors hamper functionality in all quantum network stages of state preparation and measurement, entanglement distribution, and entanglement generation execution. In this paper, we focus on the issue of quantum state transfer~\cite{PhysRevLett.78.3221}, an important process in quantum networks especially with regards to entanglement swapping process in quantum repeaters---and which offers implications for the aforementioned quantum network stages. Specifically, we concern ourselves with the interaction of errors occurring from noise and the effect they have on state transfer. 

Despite that quantum communication is well developed~\cite{gisin2007quantum, orieux2016recent, cozzolino2019high}, there is currently no large-scale quantum network testbed available. In contrast, there are several available quantum computers. Moreover, the usage of NISQ quantum computers has provided ample fertile ground to test various aspects of quantum networks ranging from quantum teleportation protocols~\cite{huang2020identification, hillmich2021exploiting}, graph state generation~\cite{mooney2021whole}, to testing quantum router and quantum repeater designs~\cite{behera2019demonstration,behera2019designing,das2021design}. As such, for our study, we will use qubits and gates in these NISQ quantum computers to simulate quantum state transfer. In particular, we investigate the interplay between gate and readout errors due to the importance of measurement readout of expectation values and the prevalent usage of noisy gates such as CNOT in most protocols. 

Our paper is organized as follows. In Sec.~\ref{sec:transfer_schemes}, we describe the set-up of our investigation using Qiskit~\cite{Qiskit}, outline the different quantum state transfer schemes we will compare throughout the paper, and address their initial performance. Then, in Sec.~\ref{sec:interplay} we present several plot results from numerical Qiskit simulations of the different schemes and try to identify the interplay between gate and readout errors. We then present an analytical point of view as well that matches the results seen in the initial numerical Qiskit simulations. In Sec.~\ref{sec:mitigation}, we touch upon the possible role error mitigation, mainly, zero-noise extrapolation (see Ref.~\cite{giurgica2020digital,temme2017error}) can serve in our analysis. Finally, in Sec.~\ref{sec:conclusion} we offer a summary of our results and propose future outlook.

\section{State Transfer Schemes} \label{sec:transfer_schemes}
    Due to the issue of diminishing quality of quantum states across large distances in a quantum network, efficient and effective state storage and transfer is crucial. Even with the use of intermediary repeaters, entanglement swapping remains a significant component, and hence the problem of state transfer remains \cite{munro2015inside}.
    
    To investigate effective quantum state transfers, one can look at how different schemes, or perhaps `resource schemes', can vary, and more intriguingly, affect the success probability of transferring an arbitrary initial state from a starting site $i$ to an end site $j$. These starting and end sites would be connected via some known/given connectivity topology that dictates the specific gate operations that are available for use, as well as the positioning of intermediary qubits facilitating the connection between the start and end site. Given such a setup, for simplicity,  we consider in this section onward, a qubit mapping to a linear chain connected topology, and we ask how certain schemes would fair in achieving the goal of successful state transfer -- especially under more noisy and error prone situations. Specifically, the physical setup that we consider is a collection of qubits in a quantum computer, which can be acted by single- and two-qubit gates, such as the transmon qubits of IBM Q, and treat them from the state-transfer's perspective. While still in the broader context of quantum networks, the state transfer is not across long-distance nodes but is across a few qubits geometrically nearby.

    With that in mind, in this paper we perform analysis on how the following four different schemes would fair when given such a task:
\begin{enumerate}
  \item SWAP: Sole, successive state swaps from site $i$ to site $j$;
  \item Teleportation: Sole, successive state teleportation (via creation of a sequence of Bell pairs) from site $i$ to site $j$;
  \item GHZ: State transfer from site $i$ to site $j$ via the creation of a GHZ as a channel;
  \item  Cluster: State transfer from site $i$ to site $j$ via the intermediate creation of a  cluster state as a resource channel, also known as the gate teleportation.
\end{enumerate}
Each of the four schemes considered offers some interesting variation.
One uses the least measurements (SWAP scheme), one uses the most measurements outcome dependent gates (GHZ scheme), and another uses a well known protocol involving Bell state measurements (teleportation scheme) while an alternative may be more suited to a particular qubit topology (cluster state resource scheme).

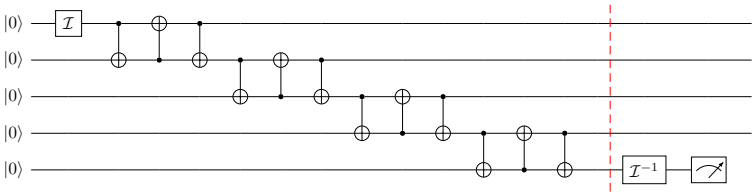
\begin{figure}[h]
    \centering
    \begin{tikzpicture}
        \node[scale=.65] {
            \begin{quantikz}[row sep={0.75cm,between origins},column sep = 0.51cm, slice style={shorten <=-0.3cm,
shorten >=-0.3cm}, thin lines]
\lstick{$\ket{0}$} &  \gate{\mathcal{I}} & [0.1cm] \ctrl{1} & \targ{}  & \ctrl{1} & \qw & \qw & \qw & \qw & \qw & \qw & \qw & \qw & \qw & \qw  & \qw & \qw & \qw
\\
\lstick{$\ket{0}$} & \qw &  \targ{}  & \ctrl{-1} &  \targ{}  & \ctrl{1} & \targ{}  & \ctrl{1} & \qw  & \qw & \qw & \qw & \qw & \qw & \qw & \qw & \qw & \qw
\\
\lstick{$\ket{0}$} & \qw & \qw & \qw & \qw &  \targ{}  & \ctrl{-1} &  \targ{}  & \ctrl{1} & \targ{}  & \ctrl{1} & \qw  & \qw & \qw & \qw & \qw & \qw & \qw
\\
\lstick{$\ket{0}$} & \qw & \qw & \qw & \qw & \qw & \qw & \qw &  \targ{}  & \ctrl{-1} &  \targ{}  &  \ctrl{1} & \targ{}  & \ctrl{1} & \qw & \qw & \qw & \qw
\\
\lstick{$\ket{0}$} & \qw & \qw & \qw & \qw & \qw & \qw & \qw & \qw & \qw &\qw &   \targ{}  & \ctrl{-1} &  \targ{}  & \qw \slice{} & \gate{\mathcal{I}^{-1}} & \meter{}
    \end{quantikz}
        };
    \end{tikzpicture}
\caption{An example five-qubit circuit schematic for the successive SWAP scheme. We note that the traditional SWAP gate remains in the three CNOT decomposed form for all simulations to study the effect of CNOT gate noise clearly.}
    \label{circ1}
\end{figure}

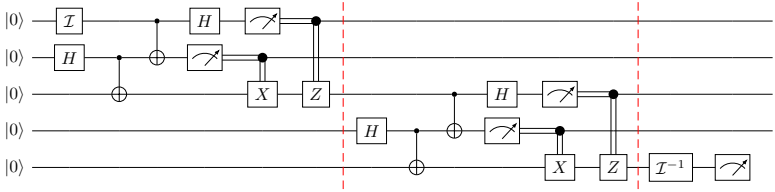
\begin{figure}[h]
    \centering
    \begin{tikzpicture}
       \node[scale=.65] {
            \begin{quantikz}[row sep={0.75cm,between origins},column sep = 0.454cm, slice style={shorten <=-0.3cm,
shorten >=-0.3cm}, thin lines]
\lstick{$\ket{0}$} &  \gate{\mathcal{I}} & [0.1cm] \qw & \ctrl{1}& \gate{H} & \meter{} & \cwbend{2} \slice{} & [0.1cm] \qw & \qw & \qw & \qw & \qw & \qw & \qw & \qw & \qw
\\
\lstick{$\ket{0}$} & \gate{H}  & \ctrl{1} &  \targ{} & \meter{} & \cwbend{1} &  \qw  & \qw & \qw & \qw & \qw & \qw & \qw & \qw & \qw & \qw
\\
\lstick{$\ket{0}$} & \qw & \targ{} & \qw & \qw & \gate{X} & \gate{Z} & \qw & \qw & \ctrl{1}& \gate{H} & \meter{} & \cwbend{2} \slice{} & [0.1cm]  \qw  & \qw & \qw
\\
\lstick{$\ket{0}$} & \qw & \qw & \qw & \qw& \qw & \qw&\gate{H}  & \ctrl{1} &  \targ{} & \meter{} & \cwbend{1} &  \qw  & \qw  & \qw & \qw
\\
\lstick{$\ket{0}$} & \qw & \qw & \qw& \qw & \qw& \qw & \qw& \targ{} & \qw & \qw & \gate{X} & \gate{Z} & [0.2cm] \gate{\mathcal{I}^{-1}} & \meter{}
\end{quantikz}
        };
    \end{tikzpicture}
  \caption{An example five-qubit circuit schematic for the successive teleportation scheme. We note that the block of gates in the dashed sections form the standard teleportation protocol utilizing Bell state creation.}
    \label{circ2}
\end{figure}

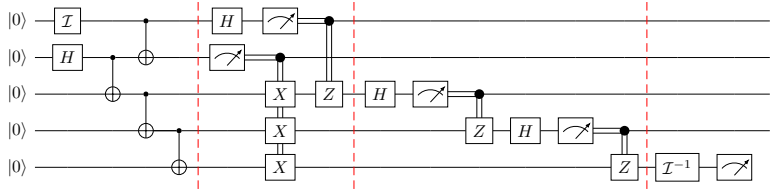
\begin{figure}[h]
    \centering
    \begin{tikzpicture}
        \node[scale=.65] {
            \begin{quantikz}[row sep={0.75cm,between origins},column sep = 0.36cm, slice style={shorten <=-0.3cm,
shorten >=-0.3cm}, thin lines]
\lstick{$\ket{0}$} &  \gate{\mathcal{I}} & [0.1cm] \qw & \ctrl{1}& \qw \slice{} & [0.1cm] \gate{H} & \meter{} & \cwbend{2} \slice{}& [0.1cm] \qw & \qw & \qw & \qw & \qw & \qw & \qw & \qw & \qw
\\
\lstick{$\ket{0}$} & \gate{H}  & \ctrl{1} &  \targ{} & \qw & \meter{} & \cwbend{3} &  \qw  & \qw & \qw & \qw & \qw & \qw & \qw & \qw & \qw & \qw
\\
\lstick{$\ket{0}$} & \qw & \targ{} & \ctrl{1} & \qw & \qw & \gate{X} & \gate{Z} & \gate{H} & \meter{} & \cwbend{1} & \qw  & \qw & \qw& \qw& \qw & \qw
\\
\lstick{$\ket{0}$} & \qw & \qw & \targ{} & \ctrl{1} \qw & \qw&\gate{X}  & \qw & \qw & \qw & \gate{Z} & \gate{H} & \meter{} &\cwbend{1} & \qw & \qw & \qw
\\
\lstick{$\ket{0}$} & \qw & \qw & \qw&  \targ{} &\qw & \gate{X} & \qw& \qw & \qw & \qw & \qw & \qw & \gate{Z} \slice{} & [0.1cm] \gate{\mathcal{I}^{-1}} & \meter{}
\end{quantikz}
        };
    \end{tikzpicture}
 \caption{An example five-qubit circuit schematic for the GHZ scheme wherein an initial GHZ state is used like a channel for state transfer from the first to last qubit.}
            \label{circ3}
\end{figure}

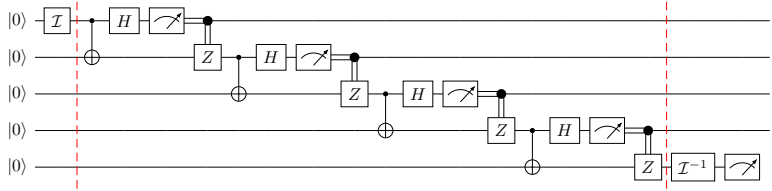
\begin{figure}[h]
    \centering
    \begin{tikzpicture}
        \node[scale=.65] {
            \begin{quantikz}[row sep={0.75cm,between origins},column sep = 0.2cm, slice style={shorten <=-0.3cm,
shorten >=-0.3cm}, thin lines]
\lstick{$\ket{0}$} &  \gate{\mathcal{I}} \slice{} & [0.1cm]  \ctrl{1}& \gate{H} & \meter{} & \cwbend{1} & \qw & \qw & \qw & \qw & \qw & \qw & \qw & \qw & \qw & \qw & \qw & \qw & \qw & \qw & \qw
\\
\lstick{$\ket{0}$} & \qw & \targ{} & \qw & \qw & \gate{Z} & \ctrl{1} & \gate{H} & \meter{} & \cwbend{1} & \qw & \qw & \qw & \qw & \qw & \qw & \qw & \qw & \qw & \qw & \qw
\\
\lstick{$\ket{0}$} & \qw & \qw & \qw & \qw & \qw & \targ{} & \qw & \qw & \gate{Z} & \ctrl{1} & \gate{H} & \meter{} & \cwbend{1} & \qw & \qw & \qw  & \qw & \qw & \qw & \qw
\\
\lstick{$\ket{0}$} & \qw & \qw & \qw & \qw & \qw & \qw & \qw & \qw & \qw & \targ{} & \qw & \qw & \gate{Z} & \ctrl{1} & \gate{H} & \meter{} & \cwbend{1} & \qw & \qw & \qw
\\
\lstick{$\ket{0}$}& \qw & \qw & \qw & \qw & \qw & \qw & \qw & \qw & \qw & \qw & \qw & \qw & \qw & \targ{} & \qw & \qw & \gate{Z} \slice{} & [0.2cm] \gate{\mathcal{I}^{-1}} & \meter{}
\end{quantikz}
        };
    \end{tikzpicture}
 \caption{An example five-qubit, simplified circuit schematic derived from using an intermediary `resource' cluster state to facilitate state transfer in the cluster scheme. We note that in the gate teleportation used in the cluster scheme, controlled-$Z$ (CZ) gate is the native gate and the remaining qubits should be initialized in $|+\rangle$, but the circuit is translated to the CNOT-based circuits, as CNOT and CZ are related by Hadamard ($H$) gates.}
    \label{circ4}
\end{figure}

In terms of evaluating an event of a successful state transfer, we sample a random initial statevector in Qiskit (approximately from the uniform Haar measure), and initialize the qubit at site $i$ (call it the first qubit w.l.o.g) in that state using some initializing gate $\mathcal{I}$ for all the schemes. We then apply the gates of a particular state transfer scheme, and finally, in the end, apply a so-called `disentangler' (which we denote by $\mathcal{I}^{-1}$) at site $j$ (or rather, the final qubit). This $\mathcal{I}^{-1}$ operation is a gate in Qiskit which undoes the unitaries that initialized the first qubit into the starting random statevector used in the beginning of the circuit. Lastly, this is followed by a measurement in the $Z$ basis to get the probability of being in our very initial $\ket{0}$ state. This probability of measuring `0' is the {\it nominal} success probability, and when there are errors and noise, the act of obtaining `0' does not necessarily mean that the final state before  $\mathcal{I}^{-1}$ is the same random state we begin with but only indicates that the protocol was successfully completed. (This is in some sense similar to the nominal success of the entanglement swap in the DLCZ protocol when a single photon is detected at the beam splitter~\cite{duan2001long,sangouard2011quantum}.) The particular implementation of the respective schemes is best illustrated with  circuit examples, and, in the order  listed above, they can be seen in Figs.~\ref{circ1}-\ref{circ4}. We note that, to highlight and contrast any scheme advantage, we use the full set of CNOTs for the successive SWAP scheme. 
    
    To start, we consider running the four schemes on Qiskit's `QASM-simulator' and applying the device noise model of `IBM Q Montreal' (simulated via the \text{FakeMontreal}() backend provider to best emulate the real device) with no circuit mapping optimization or error mitigation applied. We restrict ourselves to only the QASM-simulator in this section as we are limited in running most of the schemes (that is, all but the SWAP scheme) on a real IBM device. In particular we require the use of real-time conditional quantum gates based on measurement results (as is needed to apply the classical corrections after quantum teleportation as an example). Unfortunately, the conditional `c$\_$if()' functionality in Qiskit cannot be used on IBM's real devices as of the writing of this paper -- though we expect it may be a feature available in the future so that actual execution of the schemes considered here can be done and their results can be compared with. We could still, of course, use the principle of delayed measurements to postpone all measurements till the end, however, that introduces additional control gates that may not be directly available with the nearest-neighbor qubits and may cloud our analysis on any advantage certain schemes may have over each other. In light of this, we will mostly employ the `QASM' simulator, as that allows us to simulate the operation of real devices and to make full use of, otherwise unavailable, measurement-outcome-based conditional gates (indicated by the double lines connecting gates in Figs.~\ref{circ1}-\ref{circ4}).
    
    \subsection{Noisy simulations}
    Given such a setup, we began simulations for each scheme ranging from three qubits to thirteen qubits (in a linear geometry) on a quantum circuit, enabling all the gate- and readout- error noise from the real-device noise model. To stay consistent between the different schemes and averages, we fix the same logical qubit to device qubit error index mapping, i.e., the qubit layout, ensuring the topology mapping remains linear. The exact layout used for simulations can be seen in Appendix \ref{appendix:a}. Fixing that layout, the circuit was then transpiled using the \text{FakeMontreal} backend, and executed using the QASM-simulator with 8192 shots -- with the final expectation value averaged over thirty random initial states using the random statevector method in Qiskit.

\begin{figure}[t]
            \centering
            \includegraphics[width=.7\linewidth]{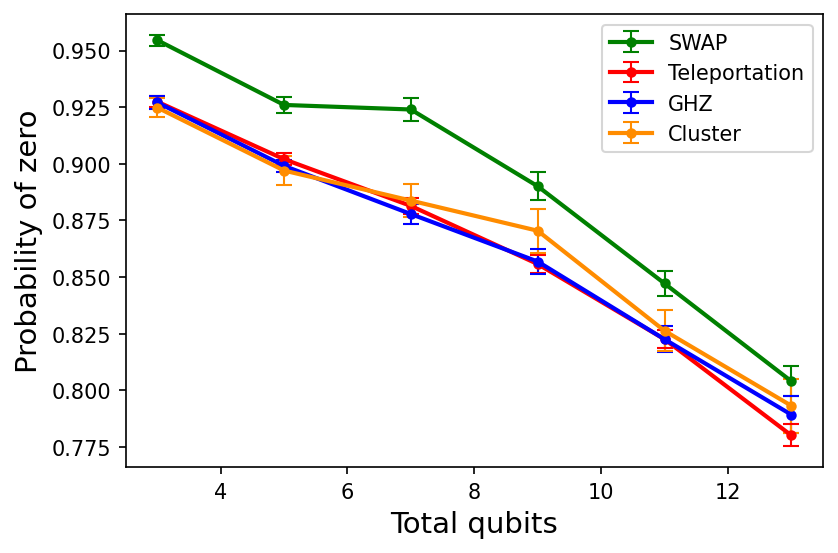}
            \caption{A plot of the expectation value of zero  (where measurement of zero indicates successful state transfer and a measurement of one, failure) varies with increasing qubit number in the different circuit schemes. The simulation used all the gate and readout errors from the device noise model of IBM \text{FakeMontreal} with 8192 shots for every data point (and each averaged over thirty random initial states). No optimization or mitigation was applied, and error bars represent the standard error in the average value.}
            \label{bar_graph1}
        \end{figure}

    We found that, overall, the SWAP scheme performed the best, followed by the cluster scheme, with the teleportation and/or GHZ scheme performing the worst as the qubits in the circuit increased (see results in Figure~\ref{bar_graph1}). This may be slightly surprising at first; the SWAP scheme utilizes the most CNOT operations (which tend to have higher gate error rates compared to single qubit gates and we ensured the SWAP scheme utilized all three CNOT gates composing a single SWAP gate) when compared to the other three schemes. Moreover, the cluster-state and teleportation schemes both utilize the same number of CNOT gates -- smaller than that of the SWAP scheme (if that is the only dominant error of concern) -- yet the cluster scheme appears to fair better than the teleportation scheme on average. Closer inspection, however, suggests that perhaps the measurement errors (or readout errors) of the device dominated, or rather, impacted the noisy simulations more so than initially anticipated. This raises a further, seemingly quite important, question of the balance between the cumulative gate errors and the cumulative readout errors in achieving an effective state transfer scheme in a noisy environment.
  
\section{Interplay of readout $\&$ gate errors} \label{sec:interplay}
\subsection{Simulations on the interplay}
\begin{figure}[h!]
\centering
\subfloat[Successful state transfer with device gate noise present only]{%
  \label{line_graph1}
  \includegraphics[clip,width=.6\linewidth]{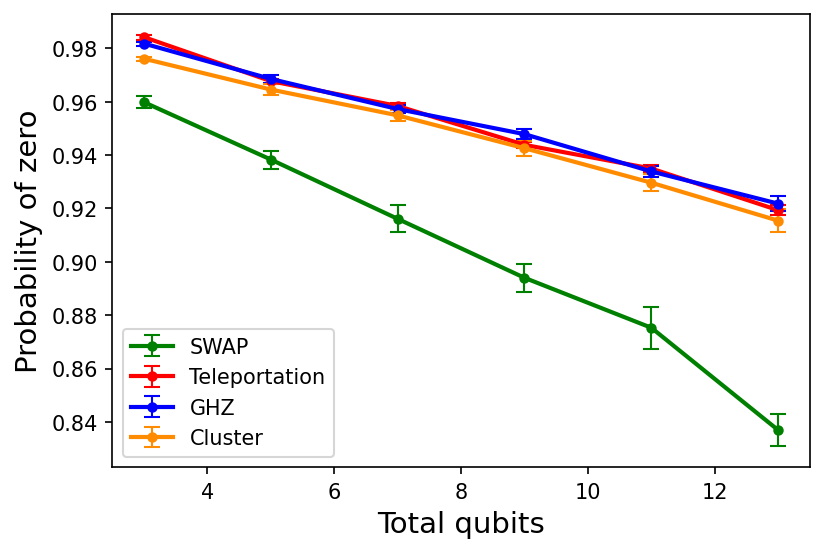}%
}\\
\subfloat[Successful state transfer with device readout errors present only]{%
  \label{line_graph2}
  \includegraphics[clip,width=.6\linewidth]{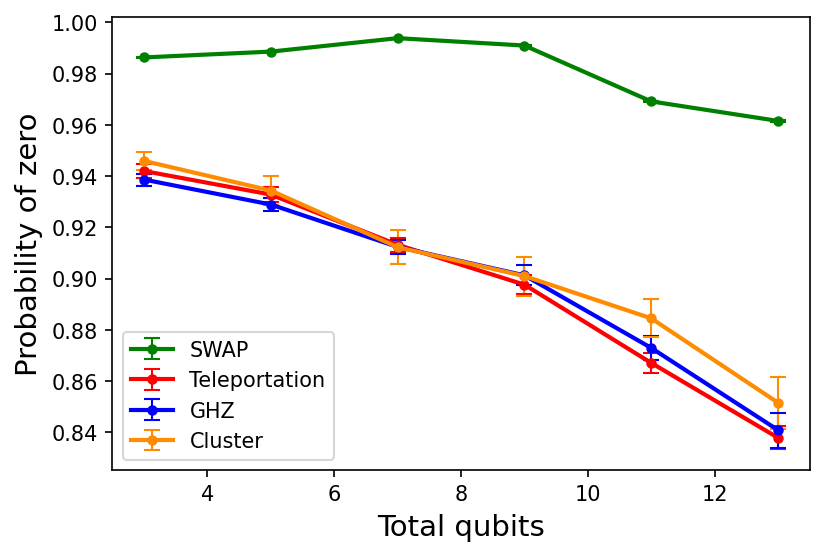}%
}
\caption{Plots (\ref{line_graph1}) and (\ref{line_graph2}) show the variation of the probability of zero when all gate errors and readout errors (in the noise model using IBM \text{FakeMontreal} as a backend) are set to zero respectively. Each data point shown was evaluated with 8192 shots and averaged over thirty randomly sampled initial states at the beginning of the circuit. No circuit optimization or mitigation was applied, and all error bars represent the standard error in the average value.}
\end{figure}

\noindent {\bf Turning off readout error}. To piece out the relative significance of the errors dominant in the previous section, we consider next the effects of having no readout errors in our noise model, namely, having ideal measurements whilst maintaining the device gate error noise. Keeping the same conditions for the simulation as previously (i.e., 8192 shots and averaging over thirty random initial states each time), the simulation results of turning off readout errors can be seen in Figure \ref{line_graph1}, which displays the relationship between the nominal successful state transfer probability (i.e., expectation of measuring zero at the final qubit) vs. the total number of qubits present in the circuit. And perhaps more predictably, we find that the SWAP scheme (the circuit with the most CNOT gates) now performs the worst, with the GHZ and teleportation schemes performing best (almost within error on average).

\smallskip\noindent {\bf Turning off gate error}.
In contrast to that, we also looked at the effects on the scheme performance when the gate errors were set to zero (that is, ideal gate operations), but with active readout errors present-- results obtained, again, from IBM \text{FakeMontreal} backend simulator. Utilizing the same topology as in previous simulations and setting any Qiskit optimization to zero, we found, interestingly and, by now expectedly, that the SWAP scheme now performs the best and the GHZ or teleportation schemes performing the worst on average (see Figure \ref{line_graph2}).

\smallskip \noindent {\bf The interplay}. To investigate the curious balance and coupled effects of having both a gate error noise model and readout error model on the different state transfer schemes, we consider how the {\it nominal} success probability varies as a function of varying levels or degree of gate noise and readout errors. To model such a simulation, we take the gate error model for the circuit basis gates to be a standard depolarizing error model in the Qiskit library (Ref.~\cite{Qiskit}) controlled by a noise level parameter $p$ (for $0 \leq p \leq 1$):
\begin{equation}
    \mathcal{E}(\rho)=\bigg(1-\frac{4p}{3}\bigg) \rho+\bigg(\frac{4p}{3} \operatorname{Tr}(\rho)\bigg)\frac{I}{2^{N}}
\end{equation}
where $\rho$ is the density matrix of the circuit, $I$ is the identity matrix, $N$ the qubit number, and {\rm Tr}($\rho)=1$ (a normalization we will mostly use).
\par Following the notation of Ref.~\cite{Qiskit}, the readout error is given via the conditional probability $P(B|A)$ which stores the probability of recording a true measurement outcome, $A$, as instead a misread outcome, $B$. For our numerical simulations we take the offset `symmetric' case where $[P(0|0), P(1|0)] = [1-P(1|0),P(1|0)] = [1 - \frac{q}{2}, \frac{q}{2}]$ and $[P(0|1), P(1|1)] = [P(0|1), 1-P(0|1)] = [q, 1-q]$. That is, we consider only a single parameter (and in that sense, we loosely say `symmetric'), $q$, that controls the probability of recording a true measurement outcome of say one, as zero for example. We note that the factor of $\frac{1}{2}$ is introduced in $[P(0|0), P(1|0)]$ to take into account that $|0\rangle$ is a lower energy state than $|1\rangle$ and hence is more likely to be measured due to decay. 
\par Given this arrangement, we define a custom noise model that is applied to all the standard basis gates in a given quantum circuit and run the simulation to find the probability of measuring zero on the final qubit in the circuit (that is, the nominal successful state transfer like defined previously). Varying the domain of $q$ and $p$ values input into our custom noise model, we can generate a surface plot of the nominal success probability as a function of $q$ and $p$. If we take, for example, say, seven qubits in each of our circuit schemes, we yield the surface plots seen in Figure \ref{num_surfaces_2} (with plots of the three and five qubits cases in Appendix \ref{appendix:b}). Each of the surface plots is composed of 1600 $(q,p)$ simulation data points to create the underlying mesh grid. Every single one of those data points used 1024 shots and whose expectation value was averaged over five random initial states (so an effective 5120 shots per data point present). 
\par Examining Figure~\ref{num_surfaces_2}, a first glance reveals a somewhat expected behavior of decreasing probability of successful state transfer with increasing depolarizing noise and/or readout error for all of the schemes. Taking a further look, however, there is a peculiar pattern that is persistent throughout all of the surface plots in which the success probability appears to be higher for high $p$ and $q$ values than when $p$ is high and $q$ is low. The pattern appears to remain persistent even when averaged over a couple of random initial states and using a large number of shots. Due to the inherent numerical nature of producing the plot, however, it can be costly (in compute) to generate a sufficiently smooth surface for the success probability to analyse more closely. As such, as an alternative, we sought to take advantage of machine learning surface regression methods from libraries such as scikit-learn SVM (support vector machines); see, e.g., Ref.~\cite{scikit-learn}. For our purposes, scikit-learn SVM has SVR (support vector regression) that would allow us to create a predicted surface regression using the raw numerical data from Figure~\ref{num_surfaces_2}. This would give us an expected surface that allows one to predict or interpolate the success probabilities, approximately, for $(q,p)$ data point values not evaluated in the raw numerical simulations of Figure~\ref{num_surfaces_2}. Such regression surfaces, for each of the state-transfer schemes (visualized with Plotly, see Ref.~\cite{plotly}) with the same seven qubits, can be seen in Figure~\ref{regression_plots_2}.

\begin{figure}[ht]
    \centering
    \subfloat[SWAP]{%
  \includegraphics[width=.44\linewidth]{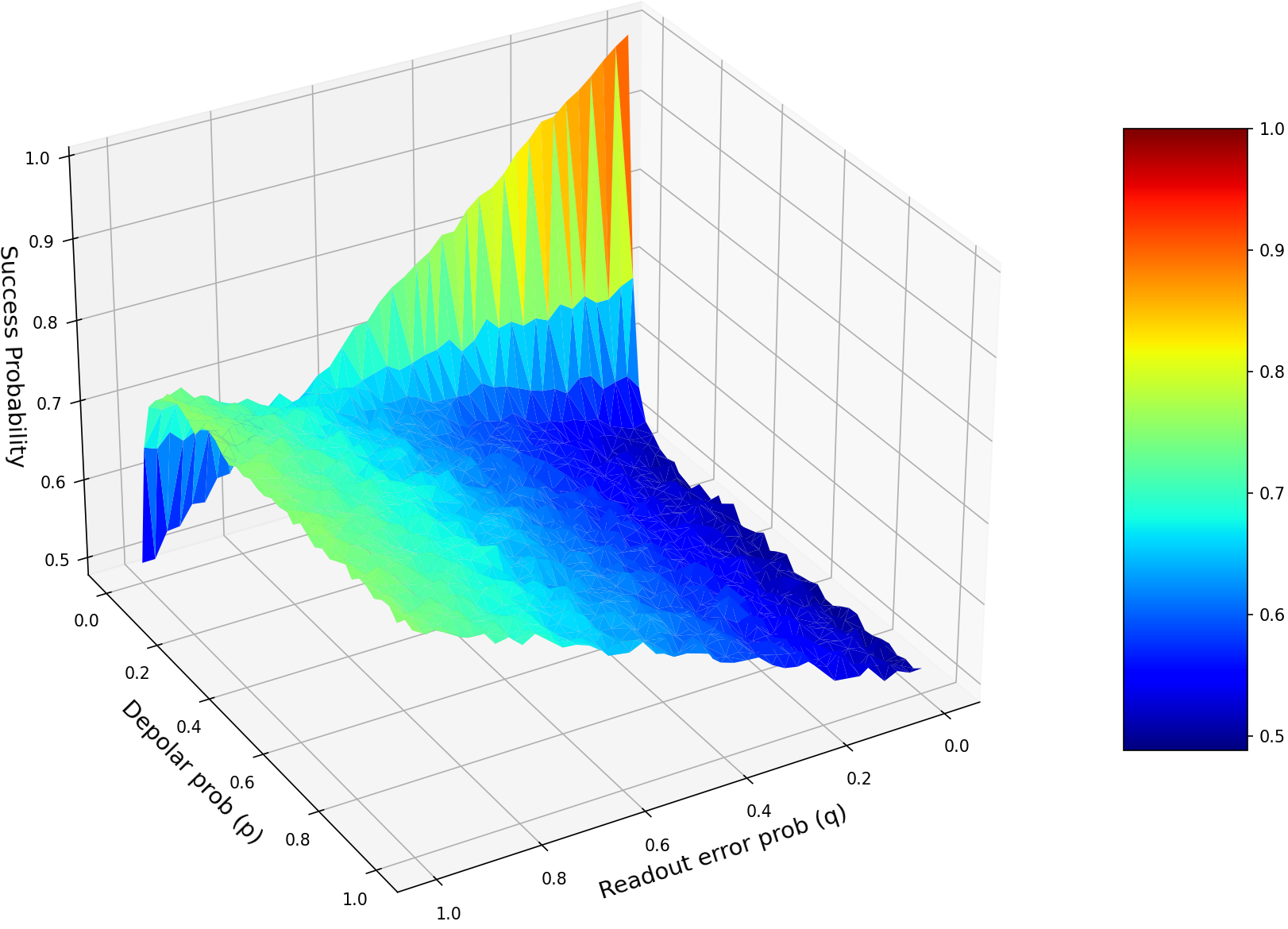}%
    }\hspace{10mm}%
    \subfloat[Teleport]{%
  \includegraphics[width=.44\linewidth]{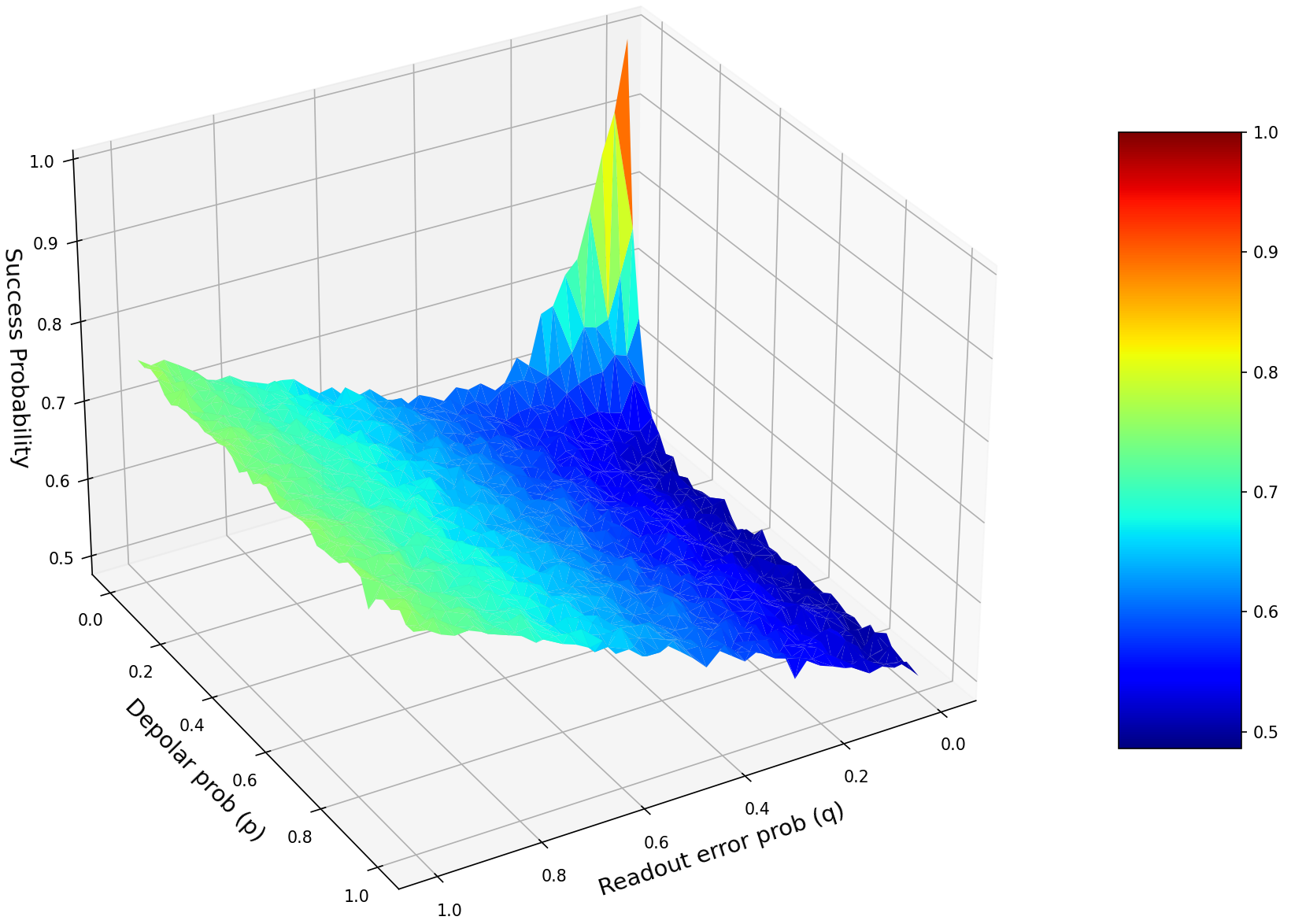}%
    }\\
    \subfloat[GHZ]{%
  \includegraphics[width=.44\linewidth]{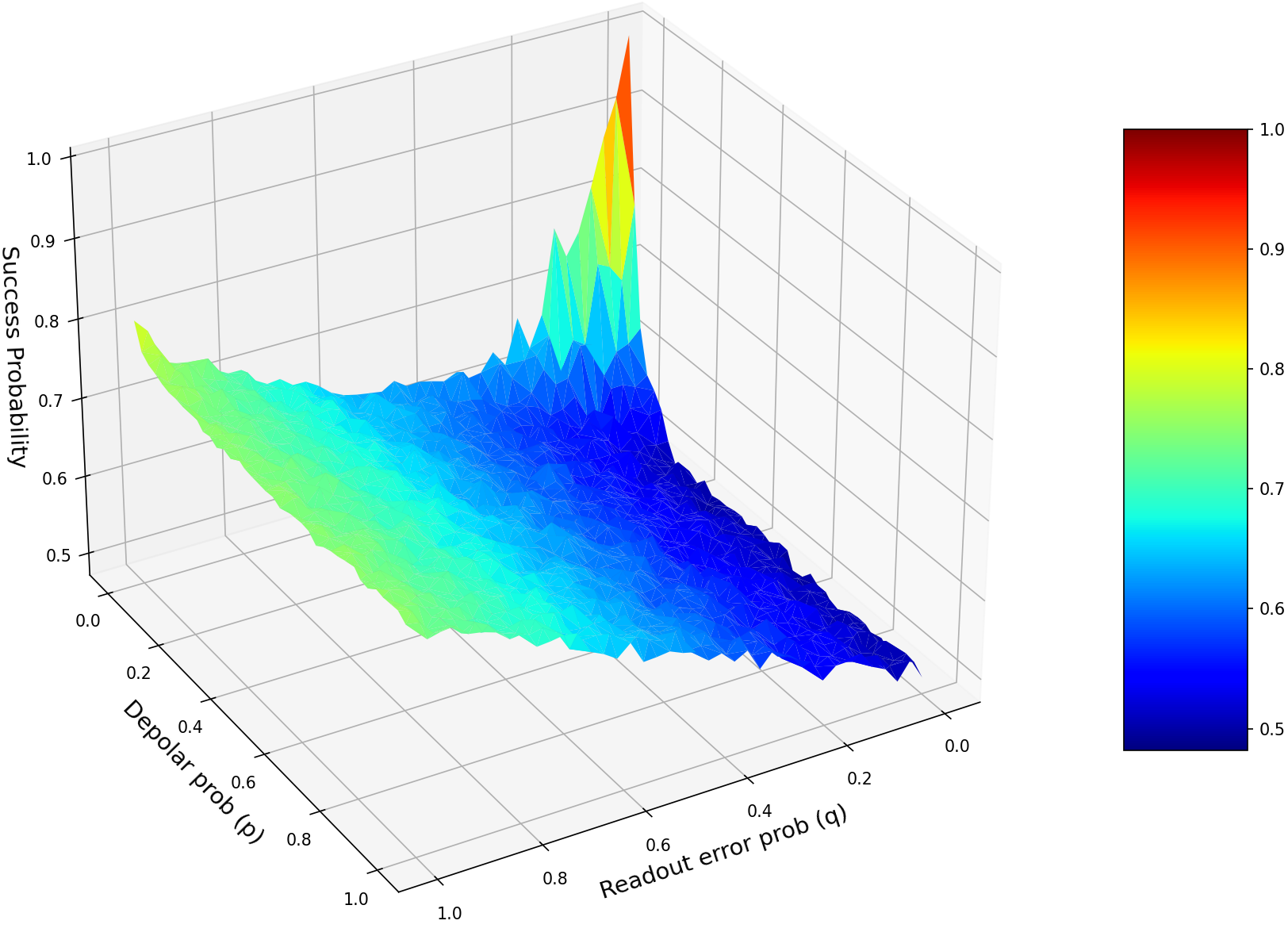}%
    }\hspace{10mm}%
    \subfloat[Cluster]{%
  \includegraphics[width=.44\linewidth]{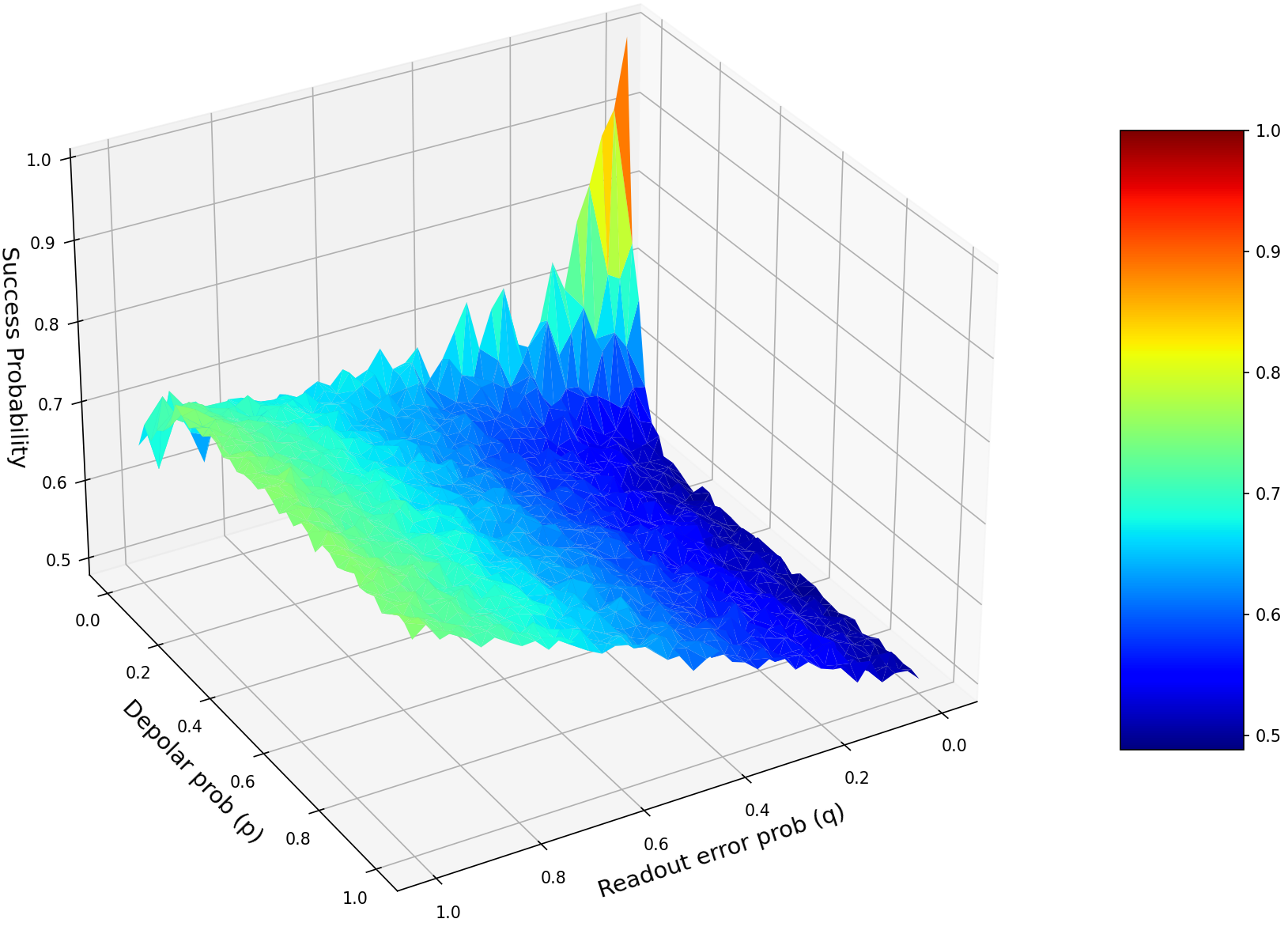}%
    }\\
    \caption{Collection of the numerical simulations composed of 1600 data points for each of the four state transfer schemes using a seven qubit circuit using 1024 shots  (and averaged over five random initial state for every single data point on the surface). The custom noise model used consisted of depolarizing gate error (controlled via parameter $p$) and a readout error model (characterized via parameter $q$), with no error mitigation applied. The z-axis of the plots represents the nominal success probability for the particular scheme of state transfer. We note that results for three and five qubits are shown in Appendix \ref{appendix:b}}
    \label{num_surfaces_2}
\end{figure}

\par These results so far implore us to consider more carefully a rather a counter-intuitive interplay between how the quantum error channel and the classical readout error model affect each other. To tease out that interplay, we turn towards an analytic understanding of the process in the next subsection. In particular, by observing that our trend of interest is present at all total qubit circuit lengths -- as low as a three qubit circuit (see Appendix \ref{appendix:b}) -- it suffices to study and analytically calculate the success probability of state transfer from a three qubit density matrix to reveal the interplay. And furthermore, we need only consider three of the four outlined schemes as the teleportation and GHZ schemes are equivalent at the three-qubit level interest.
\begin{figure}[ht]
    \centering
     \subfloat[SWAP]{%
  \includegraphics[width=.41\linewidth]{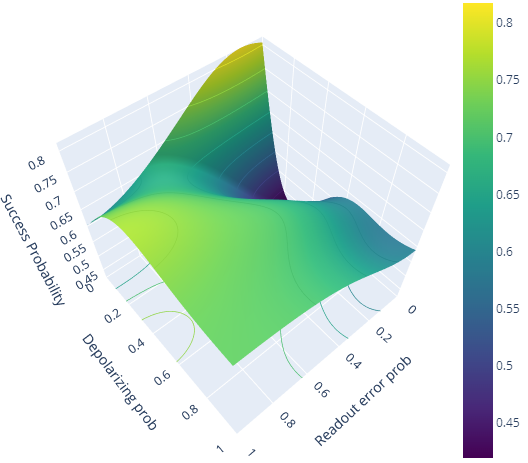}%
    }\hspace{13mm}%
       \subfloat[Teleport]{%
  \includegraphics[width=.41\linewidth]{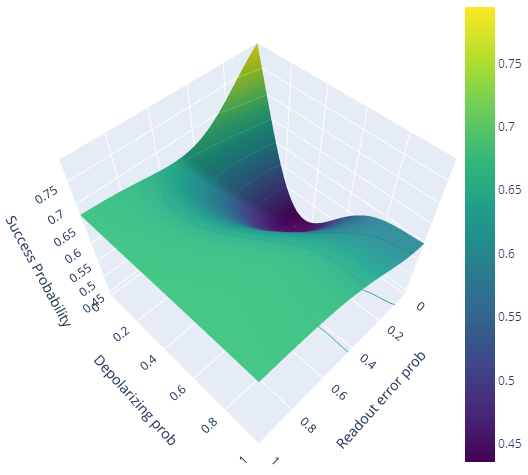}%
    }\\
       \subfloat[GHZ]{%
  \includegraphics[width=.41\linewidth]{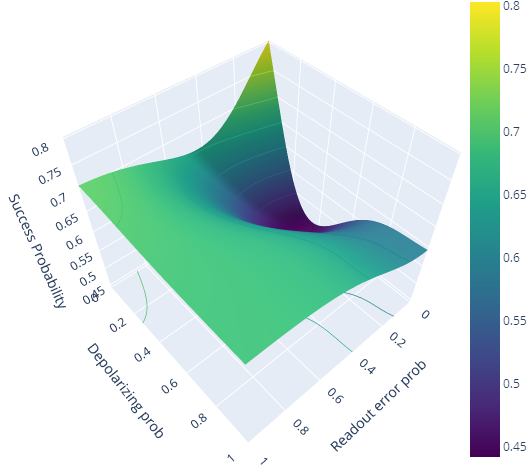}%
    }\hspace{13mm}%
       \subfloat[Cluster]{%
  \includegraphics[width=.41\linewidth]{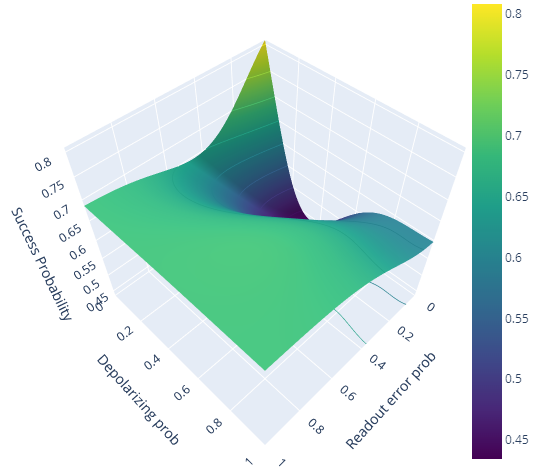}%
    }\\
    \caption{Smooth SVR (support vector regression) generated surface plots for each of the four schemes on their nominal success probability in state transfer. Similar behaviour is present like in Fig. \ref{num_surfaces_2}, with the dip in `success' probability being more starkly visible at low readout error and moderate depolarizing error.}
    \label{regression_plots_2}
\end{figure}

\subsection{Analytical results}
To start the analysis, we take an arbitrary initial density matrix, $\rho_i$, of a three-qubit system ($\text{dim}(\rho_i) = 8 \cross 8$) to be just:
\begin{equation}
\rho_{i} = \ket{\psi_i}\bra{\psi_i} = \begin{pmatrix}
\cos^2{(\theta/2)} & \frac{1}{2}e^{-i\phi}\sin{(\theta)} & 0 & \cdots & 0 \\
\frac{1}{2}e^{i\phi}\sin{(\theta)} & \sin^2{(\theta/2)} & 0 & \cdots & 0 \\
0 & 0 & \vdots & \ddots & \vdots  \\
\vdots & \vdots \\
0 & 0 & 0 & \cdots & 0
\end{pmatrix},
\end{equation}
where the components along both the row and column are listed in the order of $|000\rangle$, $|100\rangle$, $|010\rangle$, $|111\rangle$, and our initial state is given by the state $\ket{\psi_i} = \ket{\tau_i}\ket{00}$ and our arbitrary state to be transferred given by $\ket{\tau_i} = \cos{({\theta}/{2})}\ket{0} + e^{i\phi}\sin{({\theta}/{2})}\ket{1}$. We assume $\ket{\tau_i}$ can be initialized via a gate $\mathcal{I}$ (subject to gate noise) from an initial $\ket{0}$ state by applying the general $U$-gate,
\begin{equation}\label{general-UGate}
\mathcal{I} = U(\theta, \phi, 0) =\left(\begin{array}{cc}
\cos \left(\frac{\theta}{2}\right) & -\sin \left(\frac{\theta}{2}\right) \\
e^{i \phi} \sin \left(\frac{\theta}{2}\right) & e^{i\phi} \cos \left(\frac{\theta}{2}\right)
\end{array}\right),
\end{equation}
with our `disentangler' gate modelled as the inverse matrix of (\ref{general-UGate}).
\par We model our gate error channel, $\mathcal{E}$, using the standard operator-sum representation of $\mathcal{E}$ (Ref.~\cite{10.5555/1972505}) with Kraus operators $E_{i}$:
\begin{equation}
\mathcal{E}(\rho) = \sum_{i} E_{i} \rho E_{i}^{\dagger},
\end{equation}
where our primary Kraus operators used for the depolarizing channel (with probability $p$) are
\begin{equation}
E_{0}=\sqrt{1-p}I, E_{1}=\sqrt{\frac{p}{3}} X, E_{2}=\sqrt{\frac{p}{3}} Y, E_{3}=\sqrt{\frac{p}{3}}Z,
\end{equation}
and if applicable, the following for a bit flip:
\begin{equation} \label{eq:bit_flip_kraus}
E_{0}=\sqrt{\widetilde{p}}I, E_{1}=\sqrt{1-\widetilde{p}}X,
\end{equation}
and a phase-flip:
\begin{equation}\label{eq:phase_flip_kraus}
E_{0}=\sqrt{\widetilde{p}}I, E_{1}=\sqrt{1-\widetilde{p}}Z,
\end{equation}\\
channel of probability $\widetilde{p}$, with $I$ as the identity matrix and $X,Y,Z$ the standard Pauli matrices. We consider only a single-qubit version of the depolarizing channel, and merely apply a tensor product of the single-qubit depolarizing channel when adding error to two-qubits gates, such as CNOT (similar to our Qiskit numerical simulations). For all error channels, we chose to evolve the density matrix through the error channel \textit{after} applying the ideal, desired quantum gate operation to analytically model a noisy gate.
\par Next, for the final measurement readout caused error, we restrict ourselves to the classical single-qubit readout error model~\cite{maciejewski2020mitigation} (see also Ref.~\cite{chen2019detector}) on the final qubit to which the state $\ket{\tau}$ has been transferred to. Indeed, with our `disentangler' gate, we are primarily just interested in the measurement outcome `0' (`1') as our criterion for the nominal success of (failure of) state transfer. Thus, we obtain the recorded outcomes, $\mathbf{\widetilde{m}_i}$ = $(\widetilde{m}_0$,$\widetilde{m}_1)^{T}$, with our single-qubit readout error model by applying the response matrix, $\Lambda$, to the true measurement outcomes $\mathbf{m_i} = (\text{Tr}[\ket{0}\bra{0}\rho_f],\text{Tr}[\ket{1}\bra{1}\rho_f])^{T} \equiv (m_0,m_1)^{T}$ as follows:

\begin{equation}
\begin{pmatrix} \label{response_matrix}
\widetilde{m}_0\\
\widetilde{m}_1\\
\end{pmatrix}=\begin{pmatrix}
 1-q_{0} & q_{1}\\q_{0} & 1-q_{1}
\end{pmatrix}\begin{pmatrix}
m_0\\m_1
\end{pmatrix}
\end{equation}\\
where we define $q_{0} \text{ and } q_{1}$ to be conditional probabilities $P(1|0) \text{ and } P(0|1)$ respectively. To restrict the number of free parameters, we take $q_{0} = \kappa q$ (for some positive real number $\kappa$) and $q_{1} = q$ for all proceeding analyses.
\par
There exists a further issue of single-shot readout error that gives rise to accumulating readout caused error. Such errors are only of concern in the teleportation, GHZ, cluster state schemes, wherein intermediate measurement- outcome dependant errors (or just intermediate readout errors) from the ancillary qubits can lead to accumulating, incorrect application of the correcting measurement-outcome-based conditional gate. Accounting for single-shot readout error is a more difficult problem, and so for our purposes (where we primarily only care about average expectation values), we consider an alternative approximation that could account for such accumulating errors at the qualitative level. 
\par Assuming that the accumulation of such errors occur with the same probability, $q_0=\kappa q$ (or $q_1=q$ depending on random measurement outcome), we attempt to account for such intermediary errors by applying a bit-flip channel (for $X$ measurement conditional gate) and a phase-flip channel (for $Z$ measurement conditional gate). In practice, as an example, this would mean our analytical calculation for the teleportation scheme, after measurement of the first qubit as, say, zero would evolve the density matrix through a phase flip channel of probability $1-q_{0}$. In other words, the channel would model applying the identity with probability $1-q_0$ (occurs majority of the time if $q_0$ is small), and applying a $Z$ gate with probability $q_0$. This is because one would normally only need to apply the identity gate as the corrective, measurement-outcome conditional gate if the first qubit measurement outcome is one. But, a readout error model mistaking the true measurement of zero, as a one, signals that a $Z$ gate needs to be applied before proceeding (incorrectly, from an outside perspective) -- and we assume that such incorrect application, in this scenario, occurs with probability $q_0$.
\par
\begin{figure}[h!]
            \centering
            \includegraphics[width=.65\linewidth]{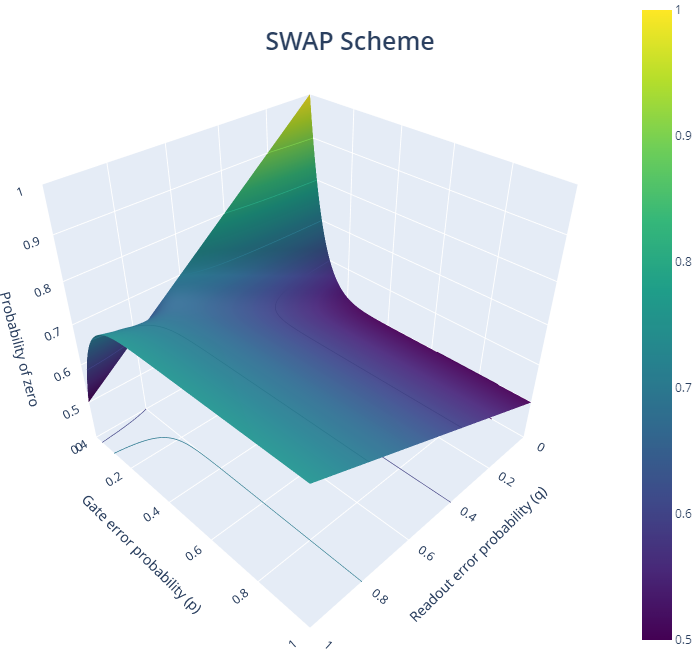}
            \caption{Nominal success probability surface plot of equation (\ref{eq:1}) (after input of equation (\ref{swap_analytic:0})) for the SWAP scheme circuit with three qubits.}
            \label{analytic_plots_1}
\end{figure}

 Given our setup so far, if we work out (via standard matrix multiplication) the evolution of $\rho_i$ through the three-qubit SWAP scheme unitaries, whilst individually applying a tensor product of the single qubit depolarizing error channel (same as done in our custom noise model in the Qiskit simulations) after every CNOT gate operation, one finds the following normalized, \emph{average} (integrated) expected measurement outcomes, $\bar{m}_{i}$, for the final qubit, as a function of the depolarizing parameter $p$:
\begin{equation}\label{swap_analytic:0}
\bar{m}_{\text{0,swap}}(p) = 1 + \sum^{12}_{j=1}\, \frac{(-1)^{j}[\mathbf{A}_{\text{swap}}]_{j}2^{(2j-1)}}{3^{(j+1)}}p^{j}
\end{equation}
where $\mathbf{A}_{\text{swap}}$ = (32, 156, 460, 915, 1296, 1344, 1032, 585, 240, 68, 12, $1)^{T}$. We note that integration over the entire Bloch sphere is performed to obtain all true average measurement outcomes in this subsection as follows:
\begin{equation}
    \bar{m}_{i} = \frac{1}{4\pi}\int_0^{2\pi}\int_0^{\pi} \text{Tr}\big[\ket{i}\bra{i}\rho_{f}(\theta,\phi)\big] \sin{\theta}\,d\theta\,d\phi.
\end{equation}
 Applying our final measurement readout error model, and using the fact that 1$-\bar{m}_{\text{0,(scheme)}} = \bar{m}_{\text{1,(scheme)}}$, we get that the nominal success probability for state transfer, that is, $\widetilde{m}_0$, is given by:
\begin{equation}  \label{eq:1}
    \widetilde{m}_{0}(q,p) = q + \bar{m}_{0}(1-(\kappa+1)q). 
\end{equation}
Plotting equation (\ref{eq:1}) using the function from equation (\ref{swap_analytic:0}) for, like our previously chosen, $\kappa=\frac{1}{2}$ (see Figure \ref{analytic_plots_1}), we begin seeing the trends observed in the previous numerical Qiskit simulations exactly. That is, for the SWAP scheme, we evidently see the same structure with the same linear drop off near the $p=0$ plane and a dip in the expectation value near low readout error and high gate error that was peculiar to us before.
\par Accordingly, following similar calculations, but now for the teleportation and cluster-state schemes (using the same depolarizing channel, but accounting for the possible accumulating, intermediary readout error, $q$, with either bit or phase flip channels), we find the following for the three-qubit teleportation (from which we can safely infer the three-qubit GHZ circuit scheme as well) and cluster-state schemes,
\begin{equation}\label{teleport+cluster_series:0}
  \widetilde{m}_{\text{0,(scheme)}} = q +  (1-(\kappa+1)q)\sum^{10}_{n=0}\sum^2_{k=0}\,[\mathbf{A}_{\text{(scheme)}}]_{nk}\frac{(-1)^{(n+k)}2^{(2n-k-1)}}{3^{(n-k+1)}}q^{k}p^{n}  
\end{equation}
with, respectively,
\begin{equation}\label{teleport_analytic:0}
\setcounter{MaxMatrixCols}{20}
(\mathbf{A}_{\text{teleport}})^T = 
\begin{pmatrix}
6 & 26 & 102 & 239 & 371 & 399 & 301 & 157 & 54 & 11 & 1 \\
4 & 36 & 147 & 359 & 581 & 651 & 511 & 277 & 99 & 21 & 2 \\
1 & 10 & 45 & 120 & 210 & 252 & 210 & 120 & 45 & 10 & 1
\end{pmatrix}
\end{equation}
and,
\begin{equation}\label{cluster_analytic:0}
(\mathbf{A}_{\text{cluster}})^T = 
\begin{pmatrix}
6 & 26 & 105 & 260 & 435 & 510 & 421 & 240 & 90 & 20 & 2 \\
4 & 40 & 180 & 480 & 840 & 1008 & 840 & 480 & 180 & 40 & 4 \\
2 & 20 & 90 & 240 & 420 & 504 & 420 & 240 & 90 & 20 & 2
\end{pmatrix}
\end{equation}

\begin{figure}[h!]
            \centering
            \includegraphics[width=.65\linewidth]{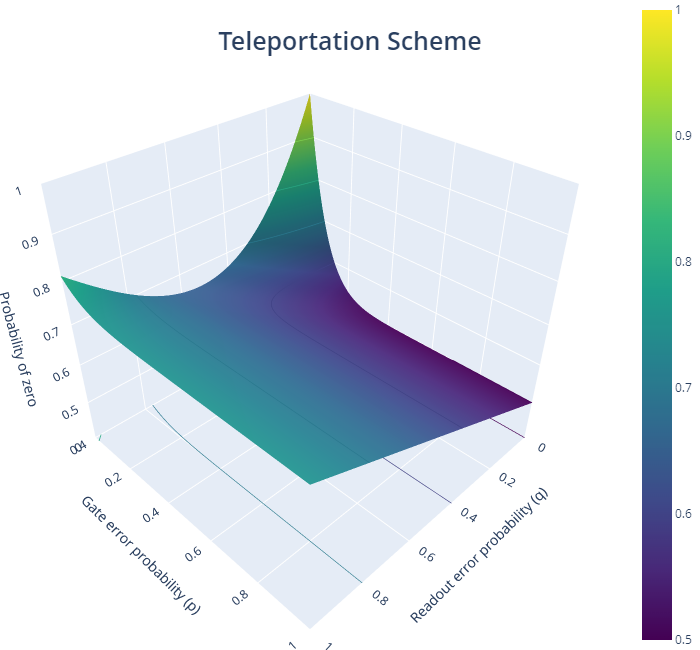}
            \caption{Nominal success probability surface plot of equation (\ref{teleport+cluster_series:0}), using (\ref{teleport_analytic:0}) for the teleportation scheme with three qubits. The plot gives insight for a three qubit GHZ scheme as well due to the circuit equivalence at the three qubit level.}
            \label{analytic_plots_2}
        \end{figure}

\begin{figure}[h!]
            \centering
            \includegraphics[width=.65\linewidth]{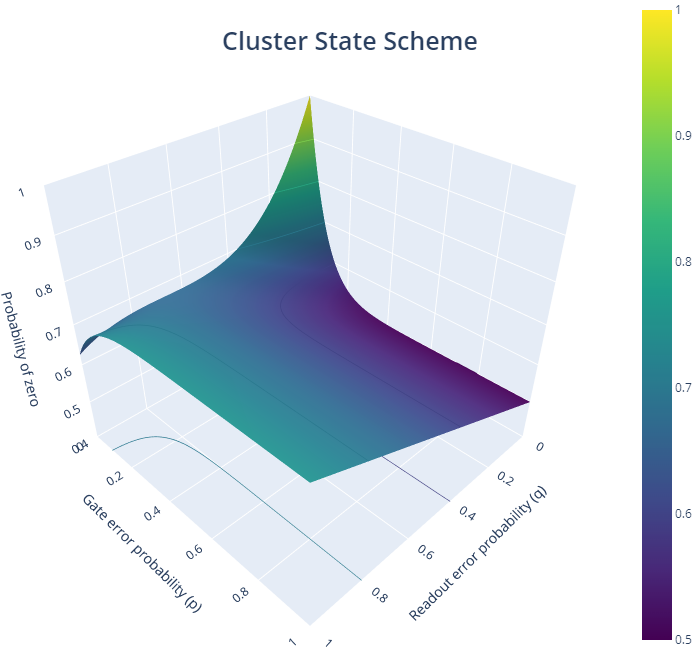}
            \caption{Nominal success probability surface plot of equation (\ref{teleport+cluster_series:0}) (using the specific input of (\ref{cluster_analytic:0})) for the cluster state scheme with three qubits.}
            \label{analytic_plots_3}
\end{figure}

We note that $\bar{m}_{0}$ for the cluster and teleportation is an average of the four possible measurement outcomes of the ancilla qubits. The plots of equation (\ref{teleport+cluster_series:0}) with coefficients from (\ref{teleport_analytic:0}) and (\ref{cluster_analytic:0}) can be seen in Figures~\ref{analytic_plots_2} and~\ref{analytic_plots_3} respectively (with $\kappa=\frac{1}{2}$ as before).

\par With our pure analytic analysis, we can now explain (and replicate) the peculiar behavior of low readout error and high gate noise resulting in the lowest successful state transfer with our analytic expressions. Indeed, from these results, we can gleam that at high readout error and high gate noise, there is so much classical readout flipping of the true measurement result that it seems to `counteract' the decrease in the measurement of zero from the gate noise. Furthermore, the asymmetry in the readout error model appears to be creating a `biasing' effect where recorded measurement of particular outcomes becomes more common at certain levels of readout error that it appears to give better results at the nominal success level. This suggests that there can exist `minima' at each readout error and gate noise produce the lowest nominal success rate and that higher values can have some sort-of `interference' effect of counteracting each other. This understanding can lead to perhaps using particular levels of readout error to one's advantage and/or utilize error mitigation techniques more effectively to `steer' where pockets (minima) of lowest, nominal success probability can appear. 

On the flip side, one could also ask if there are any similarities of such trends in the actual state fidelity as well. Now, while we cannot output the exact statevector of a qubit state on a real device, our analytic methods allow us to work out the fidelity by omitting the final `disentangler' gate unitary and final qubit measurement. Computing the density matrix evolution exactly as before, we can then compute the average state fidelity, $F$, as a function of the noise parameters $p$ and $q$ (where, again, $q$ is now a stand-in for any incorrect application of intermediary measurement dependent gate due to incorrect readout of ancilla qubits) with the commonly used formula (Ref.~\cite{hardy2001universal,bowdrey2002fidelity}):
\begin{equation}\label{fidelity_integration}
    F(q,p) = \frac{1}{4\pi} \int \bra{\tau}\rho\ket{\tau} \,d\Omega, \\
\end{equation}
where the integration is over the entire Bloch sphere again and $\ket{\tau}$ is an arbitrary initial state.
\begin{figure}[h!]
            \centering
            \includegraphics[width=.65\linewidth]{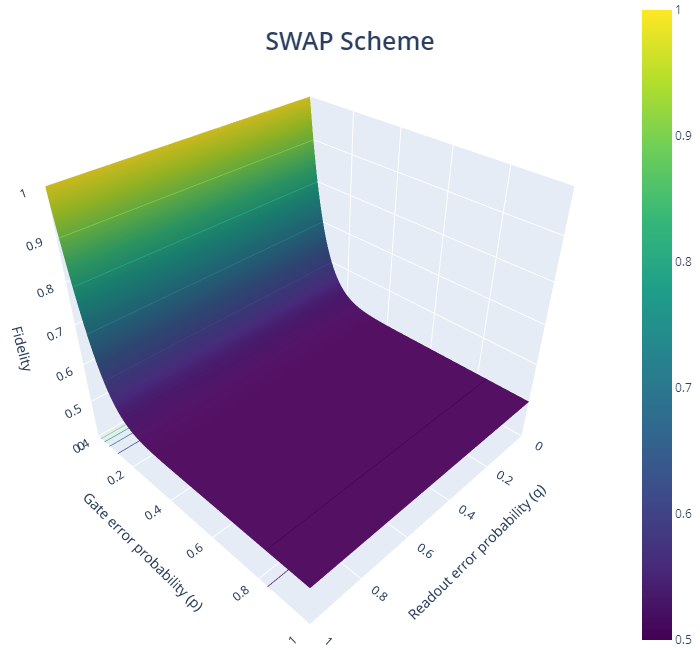}
            \caption{Final state fidelity surface plot from a three qubit SWAP scheme with varying depolarizing noise, $p$, and no intermediary readout error $q$ dependence, following equation (\ref{swap_analytic_fidelity}).}
            \label{analytic_plots_4}
        \end{figure}
\begin{figure}[h!]
            \centering
            \includegraphics[width=.65\linewidth]{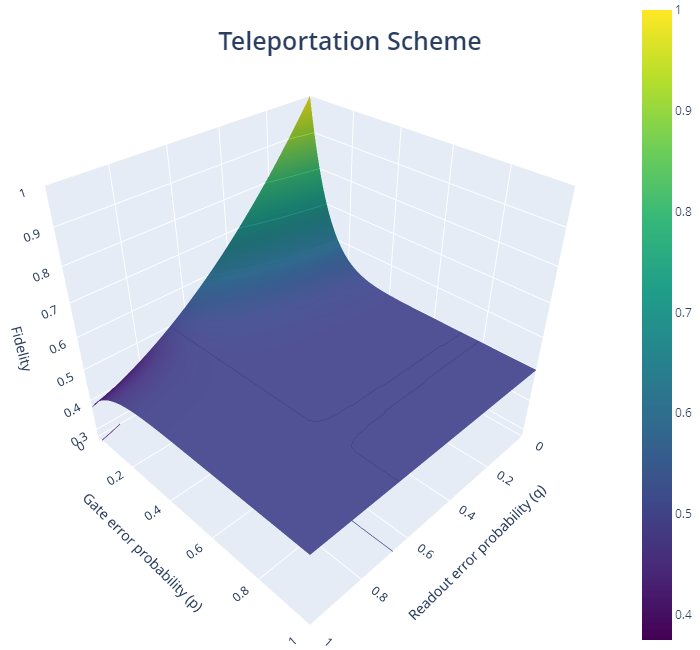}
            \caption{Final state fidelity surface plot from a three qubit teleportation scheme with varying depolarizing noise ($p$) and intermediary readout caused error, $q$, following equation (\ref{teleport+cluster_series_fidelity}) and the coefficients of (\ref{teleport_analytic_fidelity}).}
            \label{analytic_plots_5}
        \end{figure}
        
\begin{figure}[h!]
            \centering
            \includegraphics[width=.65\linewidth]{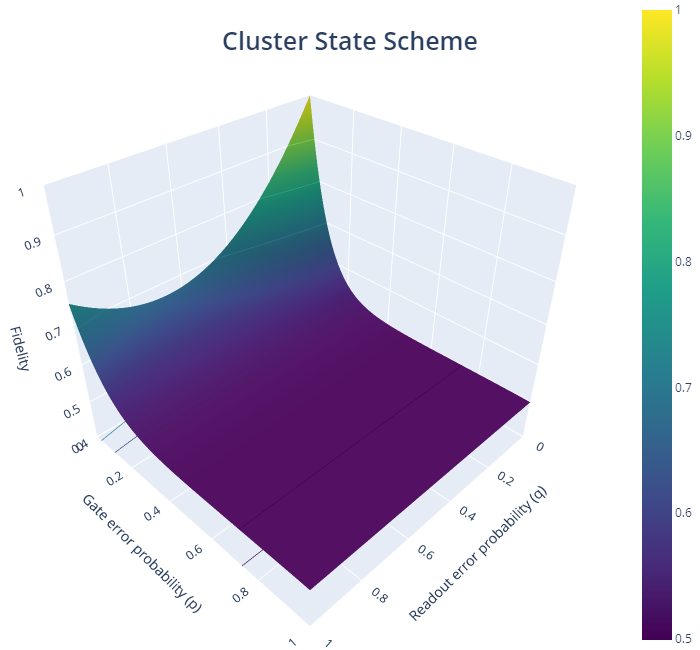}
            \caption{Final state fidelity surface plot from a three qubit teleportation scheme with varying depolarizing noise, $p$, and intermediary readout caused error, $q$, following equation (\ref{teleport+cluster_series_fidelity}) and the coefficients of (\ref{cluster_analytic_fidelity}).}
            \label{analytic_plots_6}
        \end{figure}
        
\par For the SWAP scheme, which has no measurement dependent gates, we get the average fidelity (referred to as just the fidelity in the plots and text henceforth for brevity) as purely a function of $p$:
\begin{equation}\label{swap_analytic_fidelity}
F_{\text{swap}}(p) = 1 + \sum^{11}_{j=1}\, \frac{(-1)^{j}[\mathbf{B}_{\text{swap}}]_{j}2^{(2j-1)}}{3^{(j+1)}}p^{j}
\end{equation}
where $\mathbf{B}_{\text{swap}}$ = (29, 127, 333, 582, 714, 630, 402, 183, 57, 11, 1$)^T$. For the teleportation and cluster-state schemes, the fidelity functions are now both a function of $p$ and re-contextualized $q$, and they are given by:
\begin{equation}\label{teleport+cluster_series_fidelity}
  F_{\text{(scheme)}} = \sum^{9}_{n=0}\sum^2_{k=0}\,[\mathbf{B}_{\text{(scheme)}}]_{nk}\frac{(-1)^{(n+k)}2^{(2n-k-1)}}{3^{(n-k+1)}}q^{k}p^{n}  
\end{equation}
with, respectively,
\begin{equation}\label{teleport_analytic_fidelity}
(\mathbf{B}_{\text{teleport}})^T = 
\begin{pmatrix}
6 & 23 & 79 & 160 & 211 & 188 & 113 & 44 & 10 & 1 \\
4 & 32 & 115 & 244 & 337 & 314 & 197 & 80 & 19 & 2 \\
1 & 9 & 36 & 84 & 126 & 126 & 84 & 36 & 9 & 1
\end{pmatrix}
\end{equation}
and,
\begin{equation}\label{cluster_analytic_fidelity}
(\mathbf{B}_{\text{cluster}})^T = 
\begin{pmatrix}
6 & 23 & 82 & 178 & 257 & 253 & 168 & 72 & 18 & 2 \\
4 & 36 & 144 & 336 & 504 & 504 & 336 & 144 & 36 & 4 \\
2 & 18 & 72 & 168 & 252 & 252 & 168 & 72 & 18 & 2
\end{pmatrix}
\end{equation}\\
with the respective plots of the fidelity functions shown in Figures (\ref{analytic_plots_4}-\ref{analytic_plots_6}). 

Looking at those fidelity plots, we notice that the stark dip observed previously is not quite expressly present at low readout and high depolarization noise. Though, there is some resemblance to the contours of the success probability surface plots of Figs.~\ref{analytic_plots_1}-\ref{analytic_plots_3} near the small gate error region, the high noise and high readout behavior in the fidelity surface plot is different in that the surface tends to decrease and plateau. This highlights an important note to us: the measured nominal `counts' from a real device, from a black box lens if you will, may differ significantly in behaviour to analyzing the state fidelity. If one's interest lies in just maximizing a certain expectation value, however, these results showcase what the role readout and gate error play in biasing that expectation value -- and the almost non-monotonic relationship they have to the nominal expectation value.

\section{Error mitigation possibilities} \label{sec:mitigation}

With our understanding of the role of gate and readout errors, a natural question to ask would be if we could leverage our analysis so far to mitigate errors. Given that throughout the paper we only accounted for gate errors, a technique such as zero-noise extrapolation (\cite{giurgica2020digital, temme2017error}), or ZNE for short, can be a powerful tool to mitigate noisy (from real device noise models) expectation values. The question of mitigating accumulated measurement outcome dependent errors (call it, intermediate readout resultant errors) can be more tricky. In the case of the SWAP scheme, merely inverting the response matrix, $\Lambda$, in equation (\ref{response_matrix}) would be sufficient -- as there are no measurement-outcome-based conditional gates. Unfortunately, mitigating the accumulated readout caused errors for the other schemes is not as straight-forward as that requires shot-by-shot mitigation. However, given that we are particularly dealing with mitigating average expectation values, we could attempt to apply an approximate inverse response matrix, $\Lambda^{-1}$, and assess how well it can mitigate accumulated readout error on average.
One possible way to go about mitigating real device noise could be to compare ZNE mitigated expectation values (`probability of zero' or `successful counts' depending on the context) and extrapolate the amount of accumulated readout error, on average, was needed to achieve that particular ZNE mitigated expectation value from our surface plots we have generated (on the contour of $p=0$). 

\begin{figure}[h]
            \centering
            \includegraphics[width=.7\linewidth]{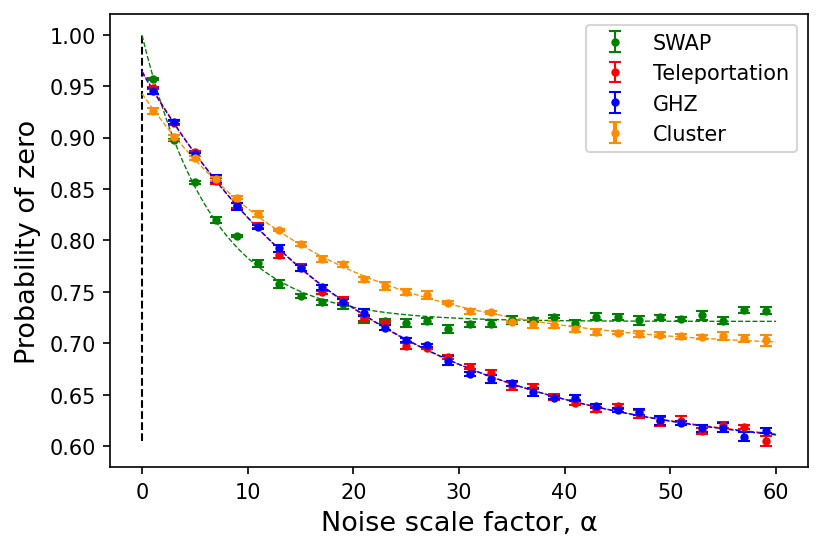}
            \caption{Three-qubit circuit scheme expectation values, $E$, for measuring zero (i.e. the probability of zero) with increasing noise scaling parameter, $\alpha$ (controlling the CNOT and $H$ gate folding in the circuits), to interpolate the zero noise limit of $E(\alpha=0)$. Noise model used for the data points is from IBM \text{FakeMontreal}, with each data point having 4098 shots and each expectation value averaged five times (with error bars representing the standard error in that average). The initial statevector sampled for this specific data is $\ket{\tau_i} = (0.79114257+0.37436334i)\ket{0} + (0.40615923+0.26264083i)\ket{1}$.}
            \label{zne_1}
        \end{figure}

With that idea in mind, a plot of how the expectation value of zero, $E$, degrades with increasing partial gateset, $G$, folding can be seen in Fig. \ref{zne_1} for a three qubit circuit. We chose to implement the gate folding following scheme outlined in Ref.~\cite{giurgica2020digital}, wherein our circuit depth, of say $D$, is scaled to $\alpha D$ after gate folding. For our circuits, we folded the Hadamard ($H$) and CNOT subset of gates and kept all else the same. From there, following Ref.~\cite{giurgica2020digital} again, we extrapolate to the zero-noise limit of $E(\alpha=0)$ (as can been seen via the vertical, dashed black line in Fig. \ref{zne_1}) by performing an exponential fit of the form $E(\alpha) \sim a e^{-b\alpha}+c$. An example dataset of the three qubit ZNE mitigated expectation values for a particular random initial state can be seen in Table \ref{tab:1}.
        
\begin{table}[h]
\begin{center}
\begin{minipage}{\textwidth}
\caption{Example dataset of expectation values of zero. The unmitigated values are the exact output of the probability of zero for a three-qubit circuit simulation (4098 shots) using IMBQ's \text{FakeMontreal} backend noise. The ZNE mitigated values are obtained from doing the extrapolation in Fig.~\ref{zne_1}. The final column shows the average readout mitigated values from inverting an approximate response matrix $\Lambda$. The initial statevector sampled for this specific dataset is $\ket{\tau_i} = (0.79114257+0.37436334i)\ket{0} + (0.40615923+0.26264083i)\ket{1}$.} \label{tab:1}
\begin{tabular*}{\textwidth}{@{\extracolsep{\fill}}lccc@{\extracolsep{\fill}}}
\toprule%
 Circuit Scheme & Unmitigated value & ZNE Mitigated value & Readout mitigated value\\
\midrule
SWAP   & 0.95583   &   0.99952 $\pm$ 0.01250 &  1.00000 $\pm$ 0.01250\\
Teleportation &  0.94436  &  0.96469 $\pm$ 0.00501 & 0.97728 $\pm$ 0.00501\\
GHZ       &  0.94802 &  0.96361 $\pm$ 0.00325 & 0.97662 $\pm$ 0.00325\\
Cluster state    &  0.92533 & 0.94260 $\pm$ 0.00301 &  0.96427 $\pm$ 0.00301\\
\botrule
\end{tabular*}
\end{minipage}
\end{center}
\end{table} 

Next, to attempt to mitigate the accumulated readout caused error, we apply $\Lambda^{-1}$ (keeping $q_0 = \frac{q}{2} \text{ and } q_1 = q$) to the ZNE mitigated values. To get a good estimate for the average $q$ in $\Lambda^{-1}$, we extrapolate its value, for a given scheme, from the respective scheme's expectation value of zero surface plot. In particular, we would determine the $q$ value present at the point where $p=0$ and the $z$ coordinate = the unmitigated expectation value. At the three-qubit level, we can straightforwardly use our analytic equations to determine such a $q$ value for $\Lambda^{-1}$. An example of the mitigation possible by this method, for a three qubit circuit, can be seen in Table~\ref{tab:1}. 

For larger qubit circuits, where specific analytic expressions may be harder to find, one could generate surface plots akin to Fig. \ref{num_surfaces_2}, but with the axis of increasing gate noise replaced, instead, with increasing ZNE noise parameter, $\alpha$, from normal device noise. The readout error axis would need to be varied as well, in some probabilistic fashion or by simply updating the device noise model each time. With sufficient data points from many randomized circuits, one could use regression methods (like used in Fig. \ref{regression_plots_2}) to generate a smooth surface plot of the expectation value varying with increasing device noise ($\alpha$) and readout error. This could allow one to extrapolate $q$ numerically, for a given expectation value, at the $\alpha=0$ limit without any analytic expression.
\par Finally, looking at Table \ref{tab:1}, we see that the ZNE mitigation works well to improve the expectation values closer to their ideal (and more so when gate noise is dominant like in the SWAP scheme), and applying an approximate inverse response matrix allows us to mitigate all noise from the SWAP scheme in particular (within some small error margin). While the same is not exactly true for the other remaining schemes, we can still report that the method of surface $q$ extrapolation can at least provide a noticeable improvement in the expectation values for all the three qubit circuit schemes. This suggests that an approximate response matrix can still have a role in mitigating accumulated readout caused errors, even if further mitigation techniques would still be required depending on the particular scheme.

\section{Conclusion} \label{sec:conclusion}
    In this work, we have seen through the various results how different state transfer schemes (a successive standard SWAP, standard teleportation, GHZ resource, and cluster state resource) fair in performing quantum state transfer from a site $i$ to site $j$ in a linear chain geometry. We have shown the unexpectedly significant role readout error can play in the choice of state transfer scheme, and illustrated the counter-intuitive interplay readout error has with gate noise in measuring nominal expectation values. We also presented analytical equations, for the three-qubit level, that are able to reproduce and explain that counter-intuitive interplay that one sees, through a black-box lens, on a real-device. Lastly, we briefly touched upon how error mitigation may enter this picture in helping to determine the better scheme performance. While we only considered gate errors and classical readout error, there are possible extensions in investigating other noise interplay than that of just depolarizing noise. Furthermore, one could extend this work to consider quantum state transfer in different network geometries and how that changes the scheme performances. Lastly, one could also investigate more sophisticated error mitigation and correction schemes that can perhaps exploit the interplay we have seen in this paper.


\section*{Acknowledgements} \label{sec:acknowledgements}    

We thank Eden Figueroa and C. R. Ramakrishnan for useful discussions. This work was supported by the National Science Foundation under  Grants No. FET-2106447 and No. PHY-1915165. 
In particular, the research on quantum state transfer was supported by FET-2106447 and the research on the effect of noise and errors and their mitigation was supported by PHY-1915165.

\section*{Declarations}

\textbf{Data Availability} The datasets generated during and/or analysed during the current study are available from the corresponding author on reasonable request

\begin{appendices}
\section{Device connectivity}
\label{appendix:a}
\begin{figure}[h!]
 \centering          \includegraphics[width=.7\linewidth]{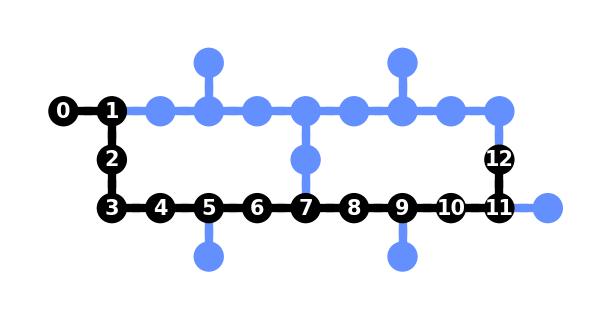}         \caption{Linear chain layout of IBM Q Montreal used for all simulations. Specifically, the custom qubit mapping to device qubit mapping is given by [0, 1, 2, 3, 4, 5, 6, 7, 8, 9, 10, 11, 12] = [0, 1, 2, 3, 5, 8, 11, 14, 16, 19, 22, 25, 24], with the latter being the physical qubit labeling on the real device.}
 \label{fig:connectivity}
\end{figure}

We show in Fig.~\ref{fig:connectivity} the layout of IBM Q Montreal whose device noise model was used in our simulations. The numbers marked in the diagram show the linear chain that we used in comparing different state transfer schemes.

\section{Success probability for 3 and 5 qubit circuits}
\label{appendix:b}
\begin{figure}[h!]
        \centering
         \subfloat[SWAP]{%
 \includegraphics[width=.35\columnwidth]{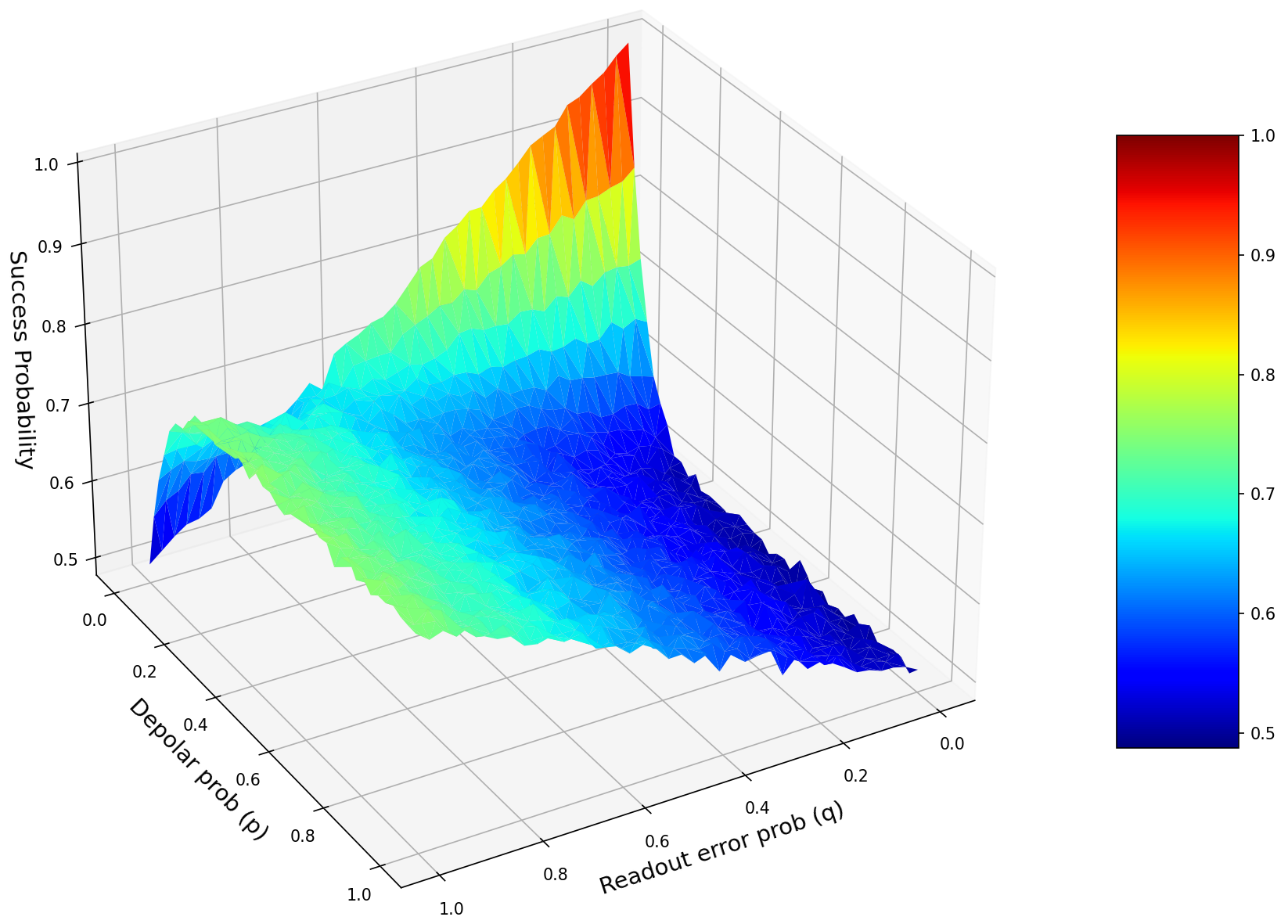}%
    }\hspace{10mm}%
    \subfloat[Teleport]{%
  \includegraphics[width=.35\columnwidth]{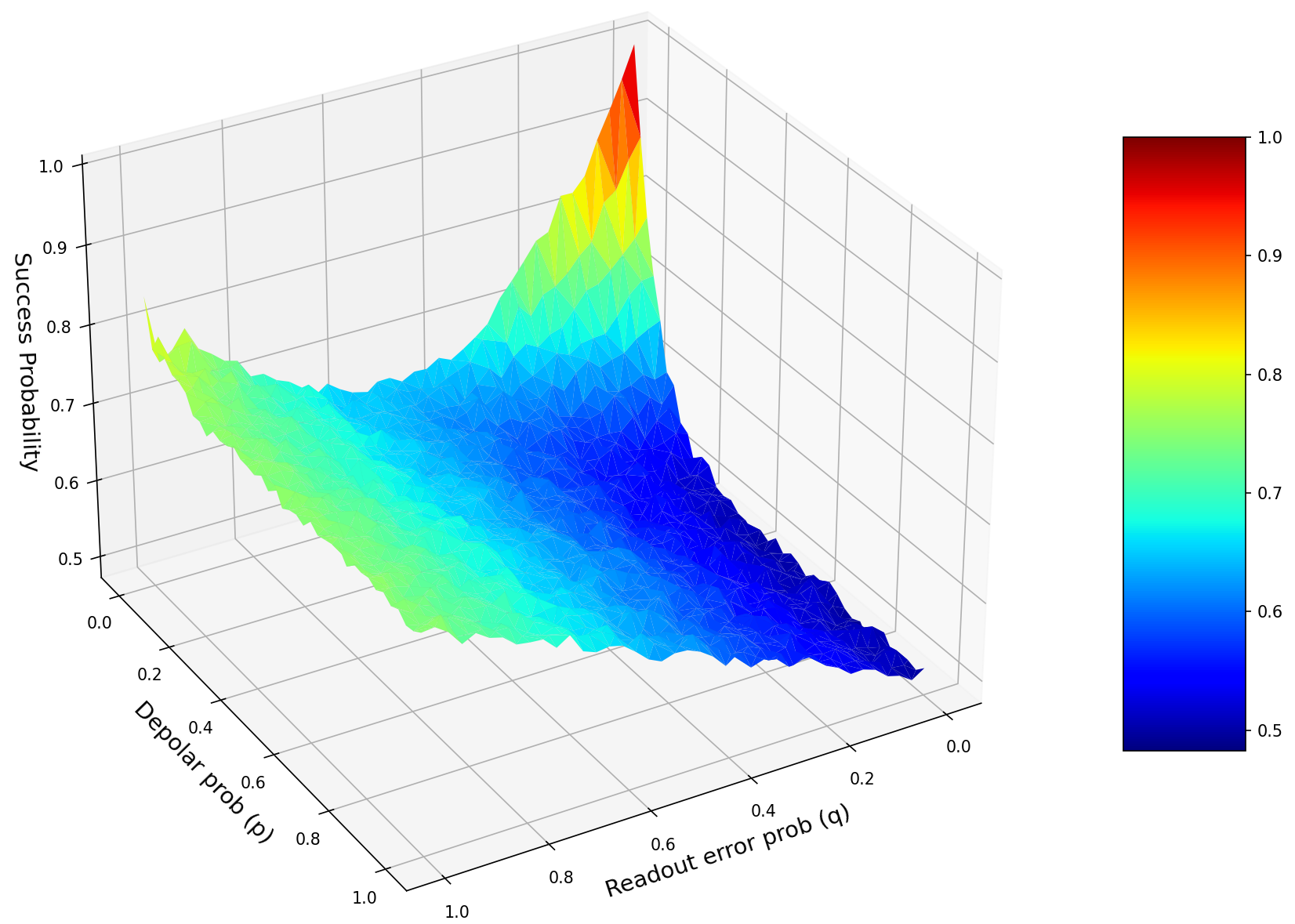}%
    }\\
    \subfloat[GHZ]{%
 \includegraphics[width=.35\columnwidth]{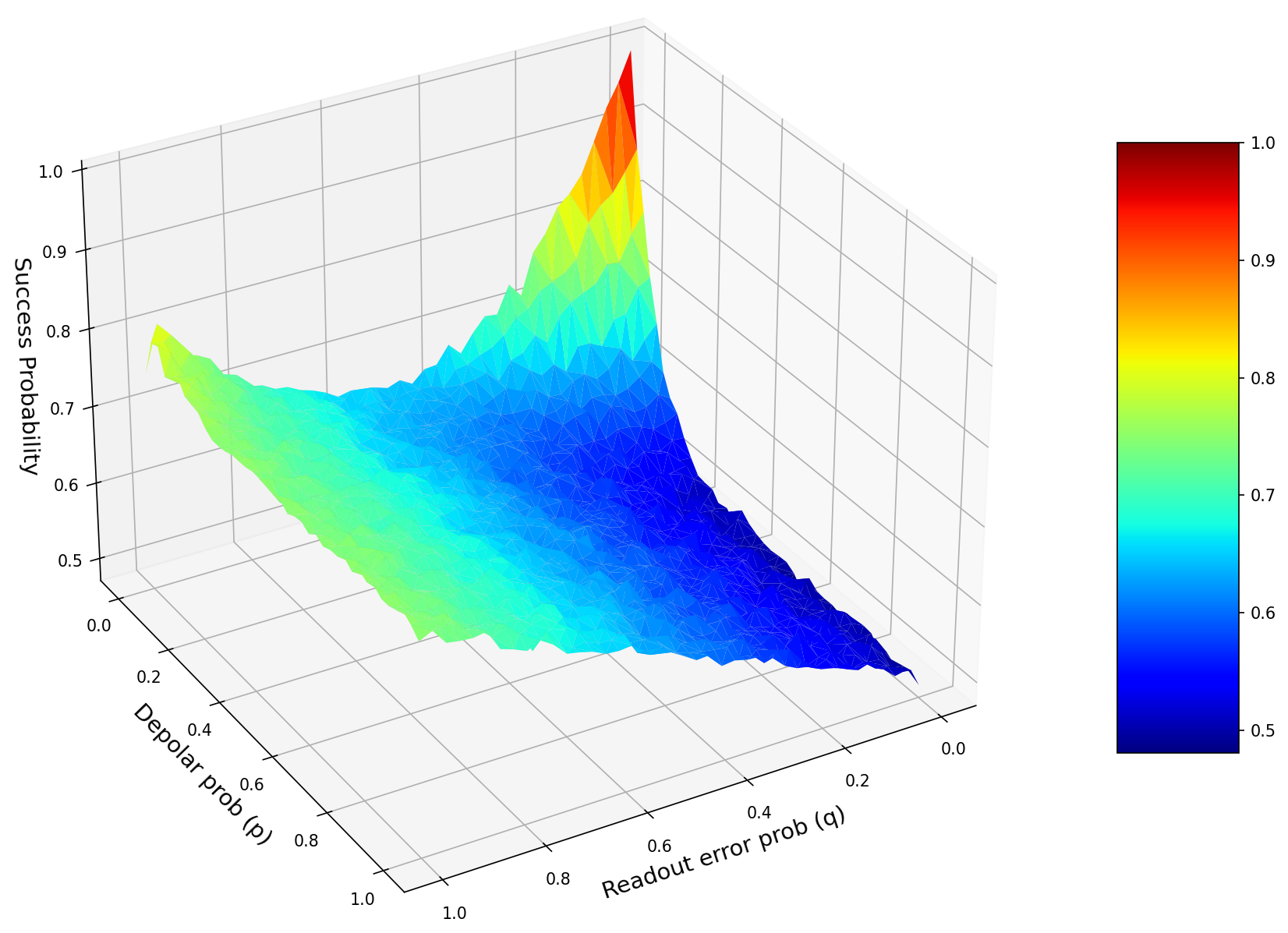}%
    }\hspace{10mm}%
    \subfloat[Cluster]{%
 \includegraphics[width=.35\columnwidth]{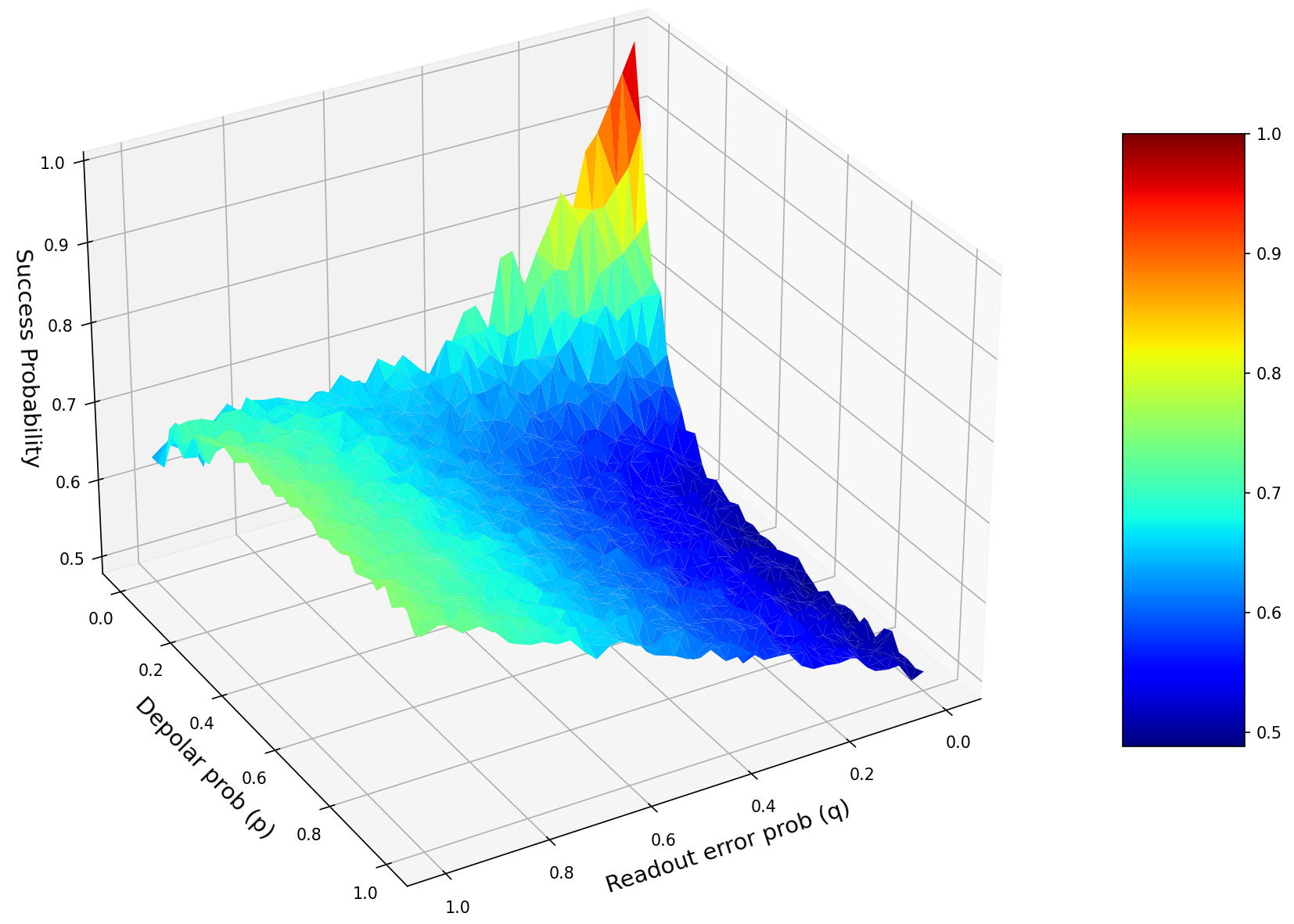}%
    }\\
    \caption{\label{fig:3qubit_surfaces}Three qubit surfaces composed of 1600 data points for each of the four state transfer schemes using 1024 shots (and averaged over five random initial state for every single data point on the surface). The custom noise model used consisted of depolarizing gate error (controlled via parameter $p$) and a readout error model (characterized via parameter $q$), with no error mitigation applied. The z-axis of the plots represents the nominal success probability for the particular scheme of state transfer.}
\end{figure}

\begin{figure}[h!]
        \centering
        \subfloat[SWAP]{%
 \includegraphics[width=.35\columnwidth]{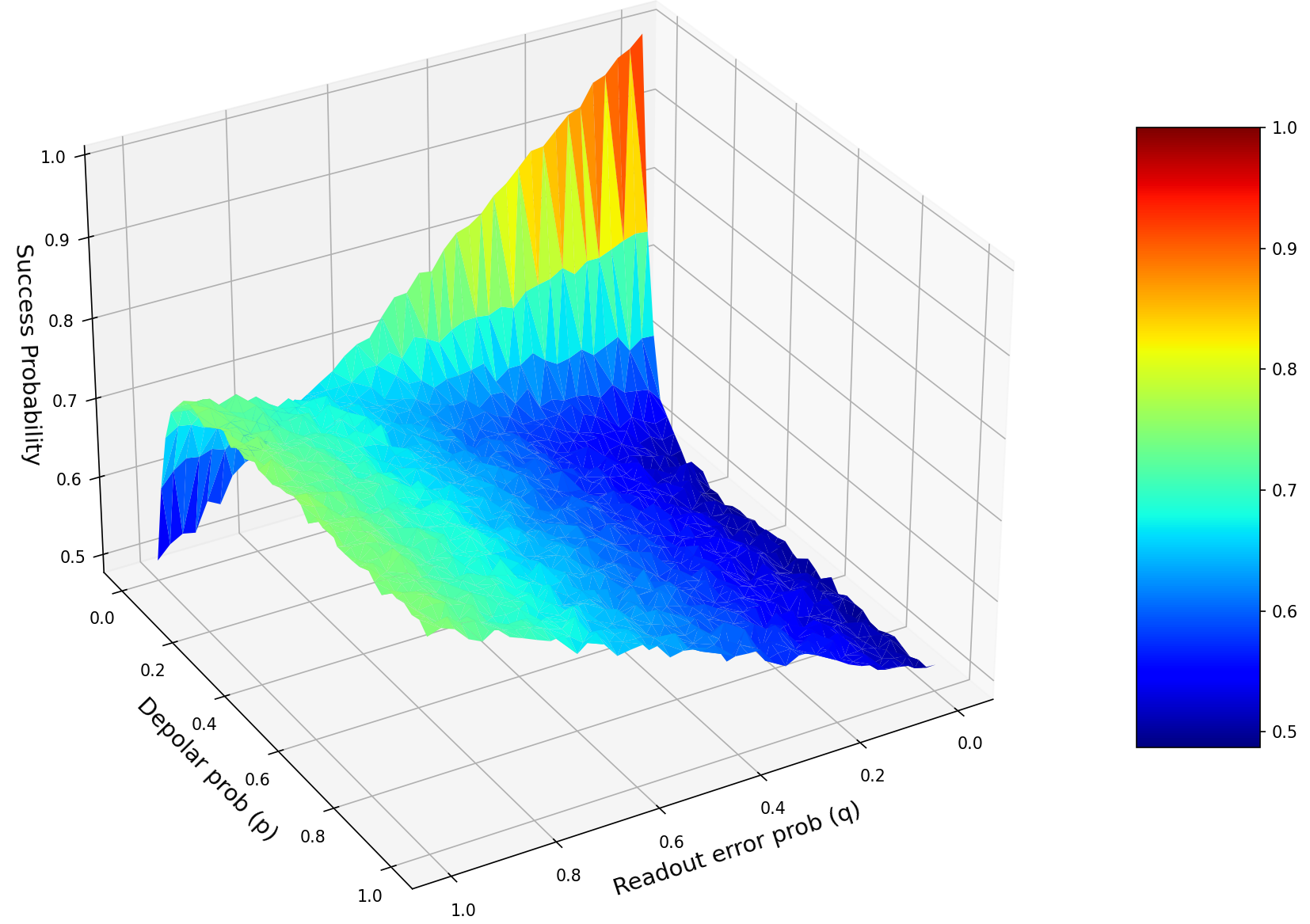}%
    }\hspace{10mm}%
    \subfloat[Teleport]{%
  \includegraphics[width=.35\columnwidth]{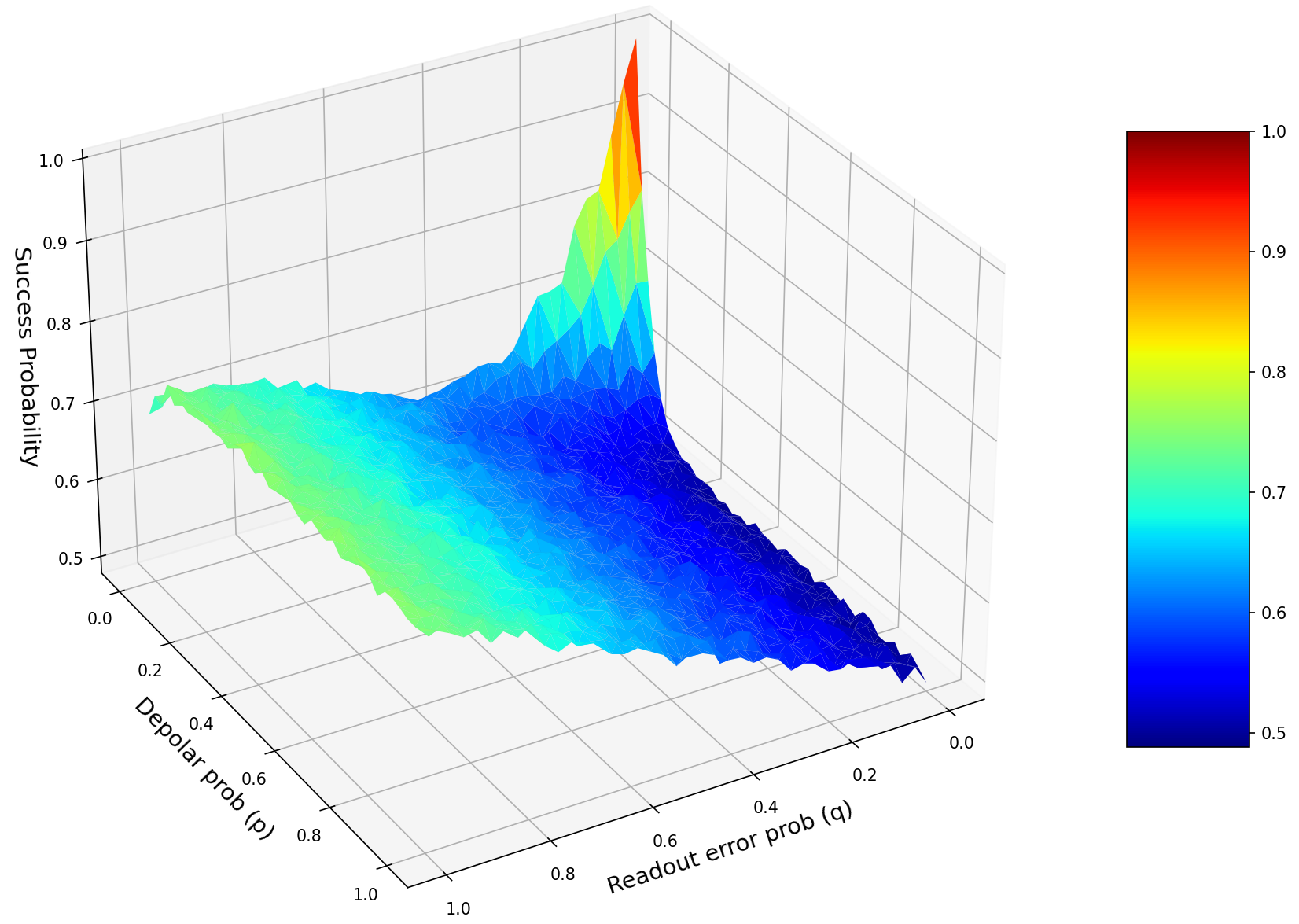}%
    }\\
    \subfloat[GHZ]{%
 \includegraphics[width=.35\columnwidth]{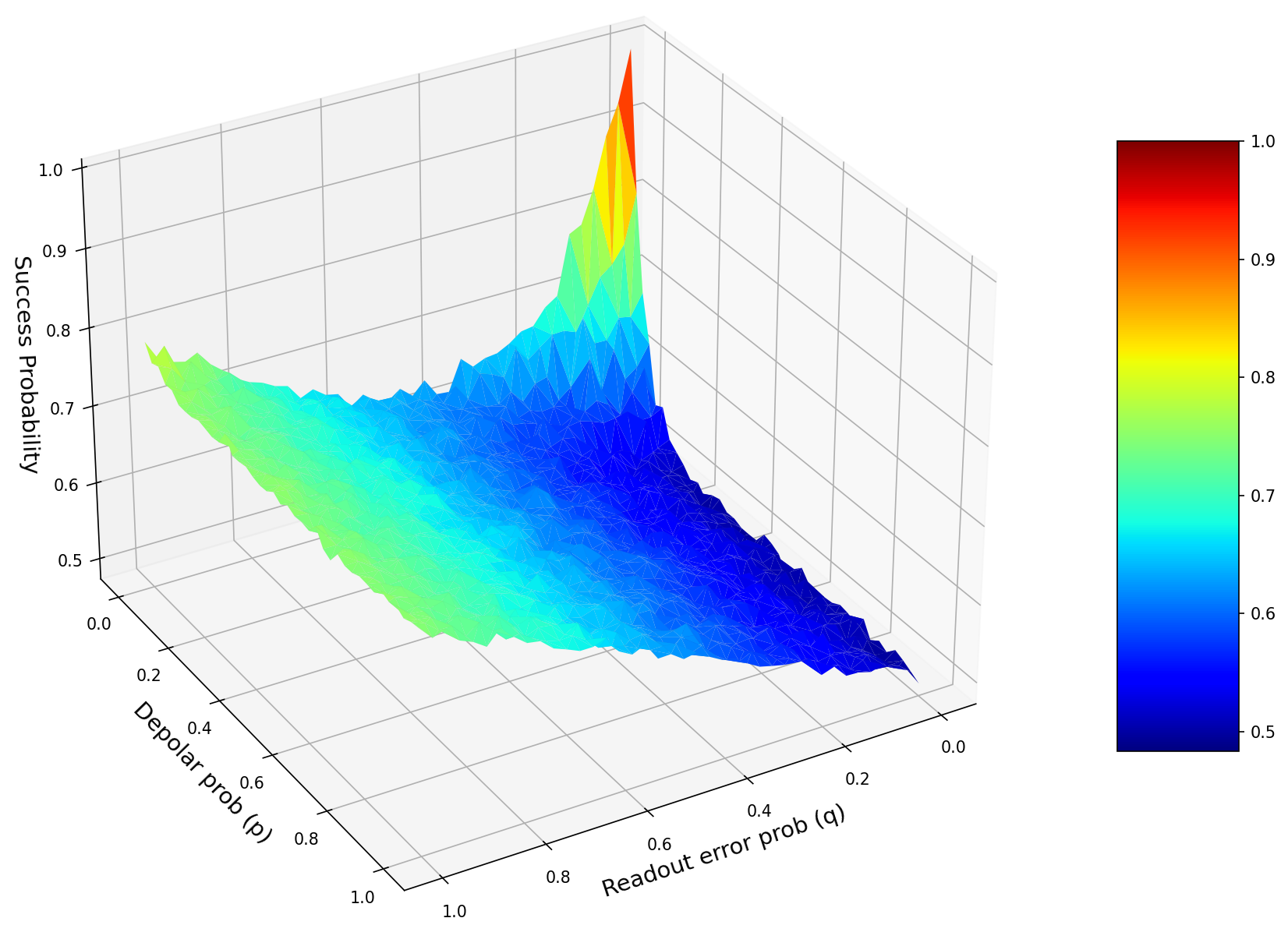}%
    }\hspace{10mm}%
    \subfloat[Cluster]{%
 \includegraphics[width=.35\columnwidth]{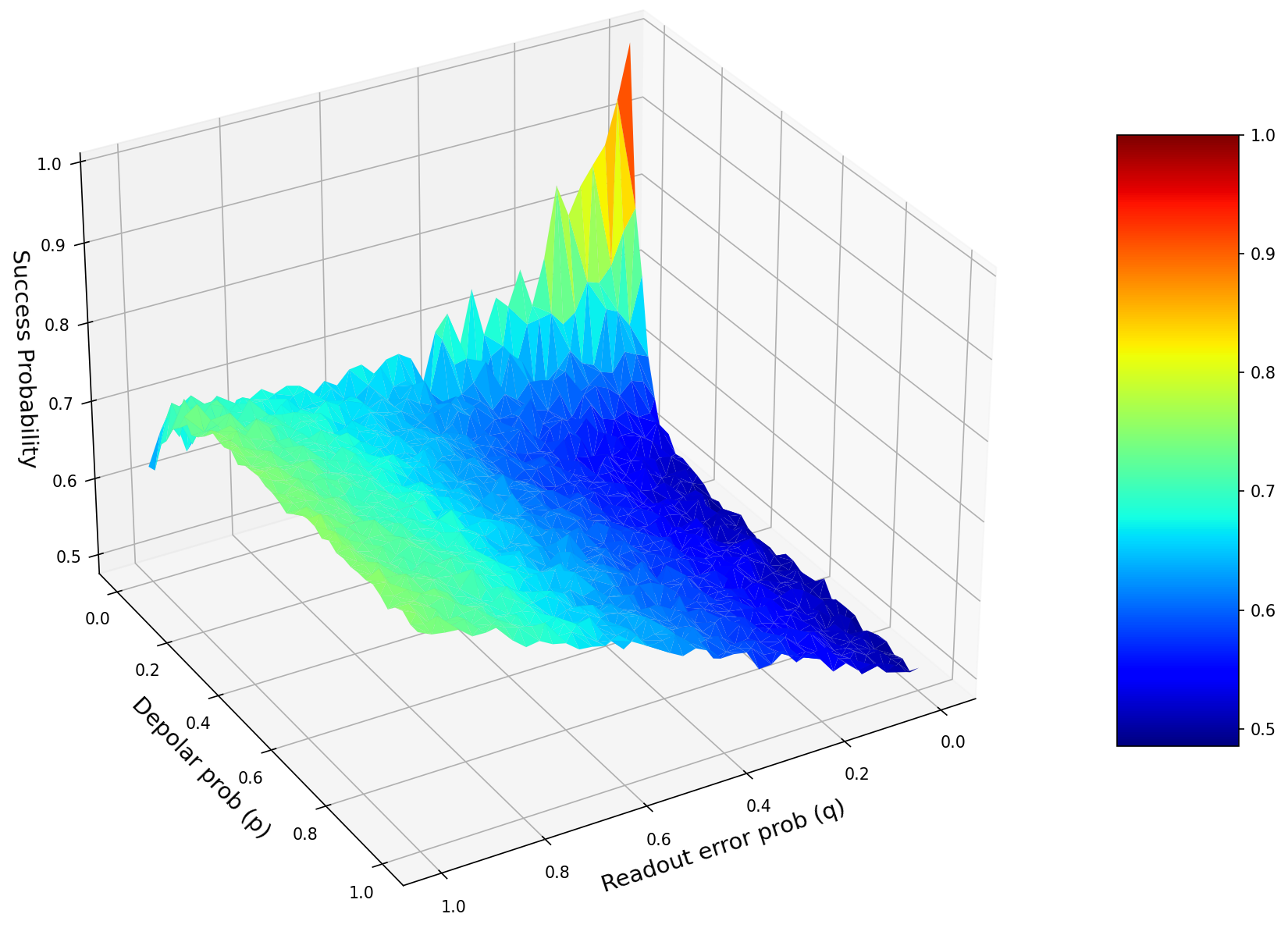}%
    }%
    \caption{\label{fig:5qubit_surfaces}Five qubit surfaces composed of 1600 data points for each of the four state transfer schemes using 1024 shots (and averaged over five random initial state for every single data point on the surface). The custom noise model used consisted of depolarizing gate error (controlled via parameter $p$) and a readout error model (characterized via parameter $q$), with no error mitigation applied. The z-axis of the plots represents the nominal success probability for the particular scheme of state transfer.}
    \end{figure}

In Fig.~\ref{num_surfaces_2}, we showed the success probability of different state transfer schemes using 7-qubit  circuits; in Figs.~\ref{fig:3qubit_surfaces}-\ref{fig:5qubit_surfaces}, we show the corresponding results with 3 and 5 qubits respectively.


\section{Hellinger fidelity for 3 qubit scheme}
As an aside, we could try to consider another measure for `successful' state transfer in Qiskit such as the so-called Hellinger fidelity, $F_H$, \cite{Qiskit} defined as:
\begin{equation}
    F_H = \left(1-h^2\right)^2
\end{equation}
where $h$ is the Hellinger distance. This distance $h$ generally tells you the `closeness' of two probability distributions (in our case, `counts' distribution) given some observations (see Ref.~\cite{luo2004informational} for more details). We note that this $F_H$ quantity reduces to the quantum state fidelity for diagonal density matrices (i.e. the classical fidelity) \cite{Qiskit}.
\par
\begin{figure}[h!]
    \centering
    \subfloat[SWAP]{%
  \includegraphics[width=.35\linewidth]{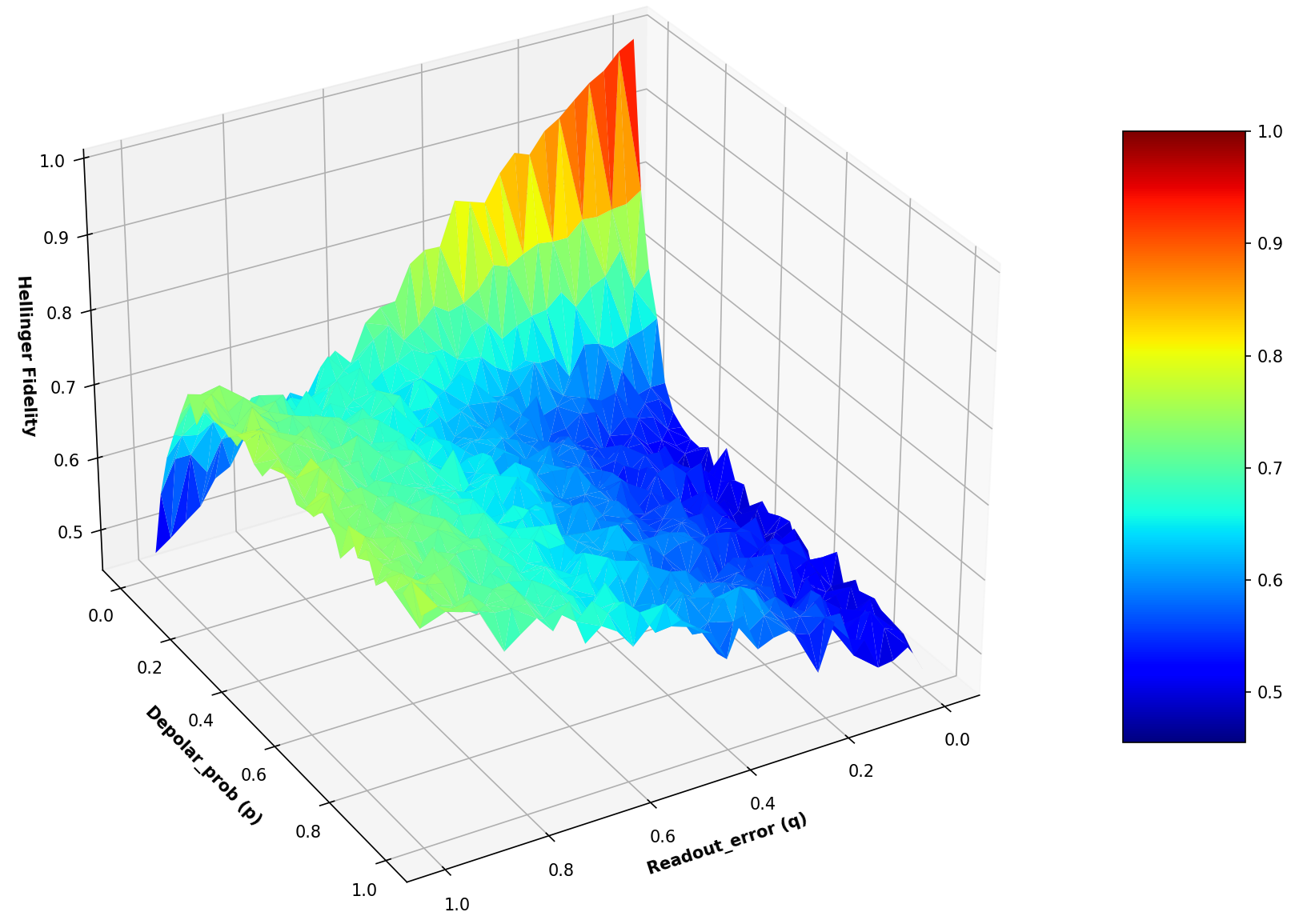}%
    }\hspace{10mm}%
    \subfloat[Teleport]{%
  \includegraphics[width=.35\linewidth]{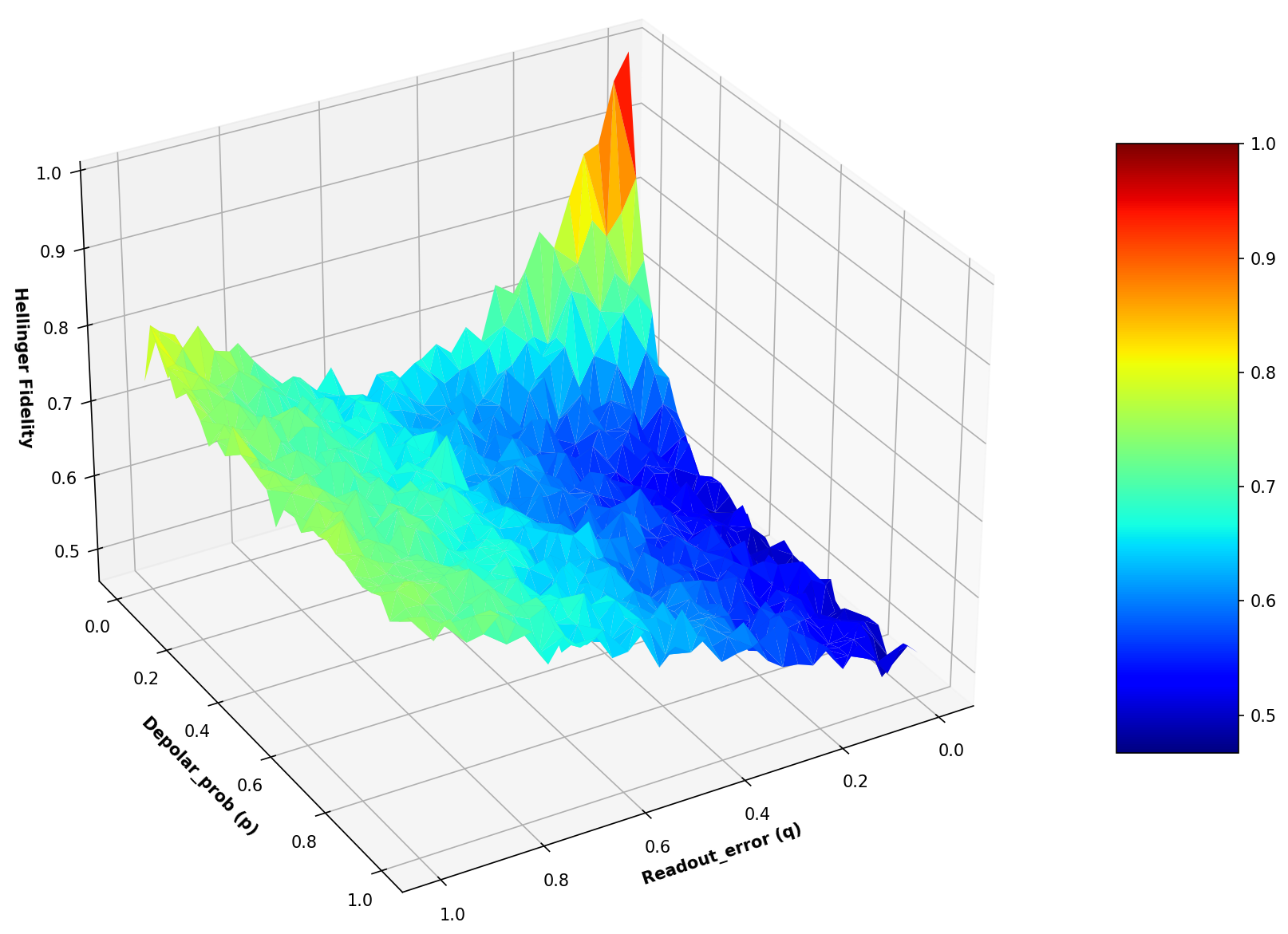}%
    }\\
    \subfloat[GHZ]{%
  \includegraphics[width=.35\linewidth]{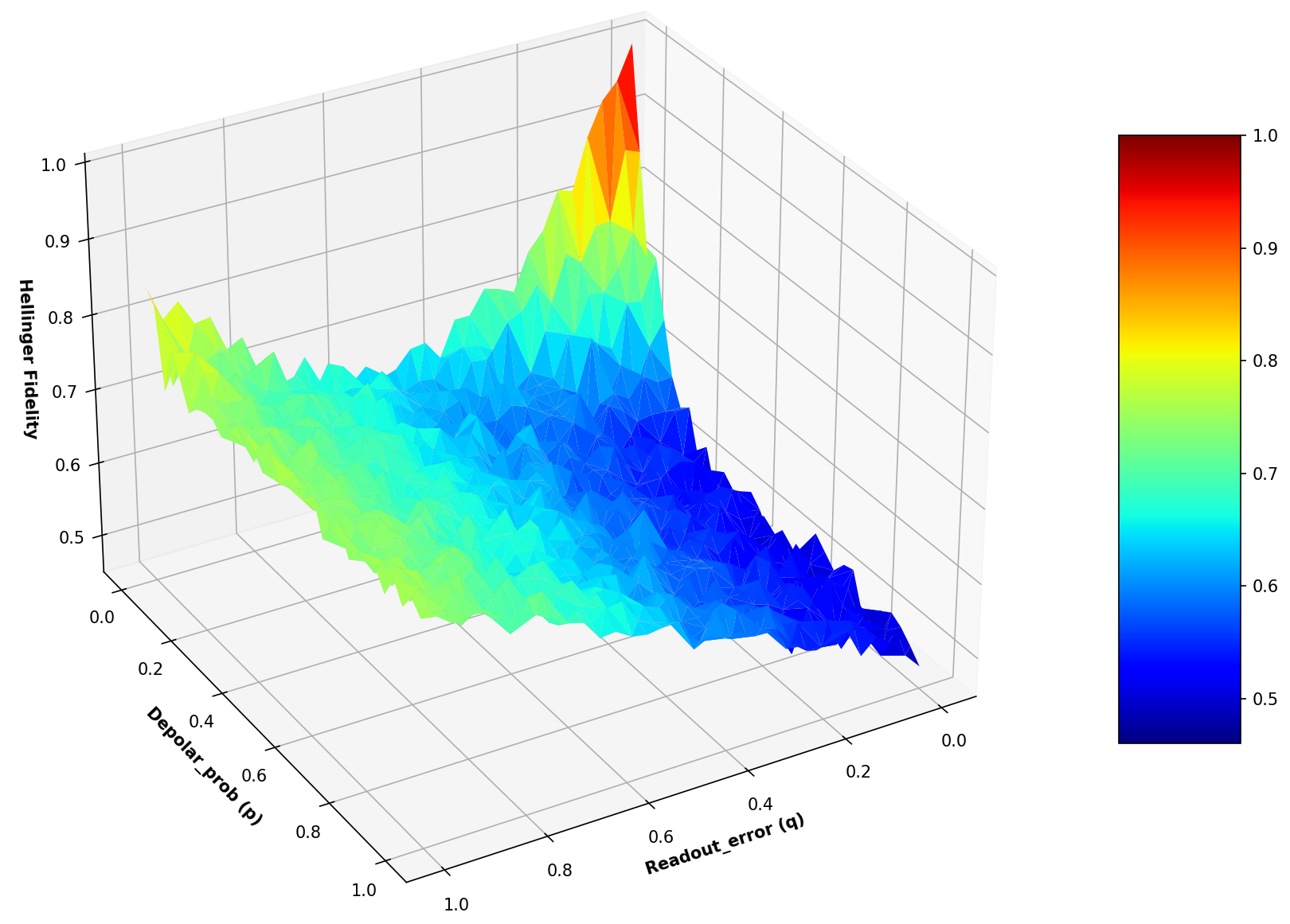}%
    }\hspace{10mm}%
    \subfloat[Cluster]{%
  \includegraphics[width=.35\linewidth]{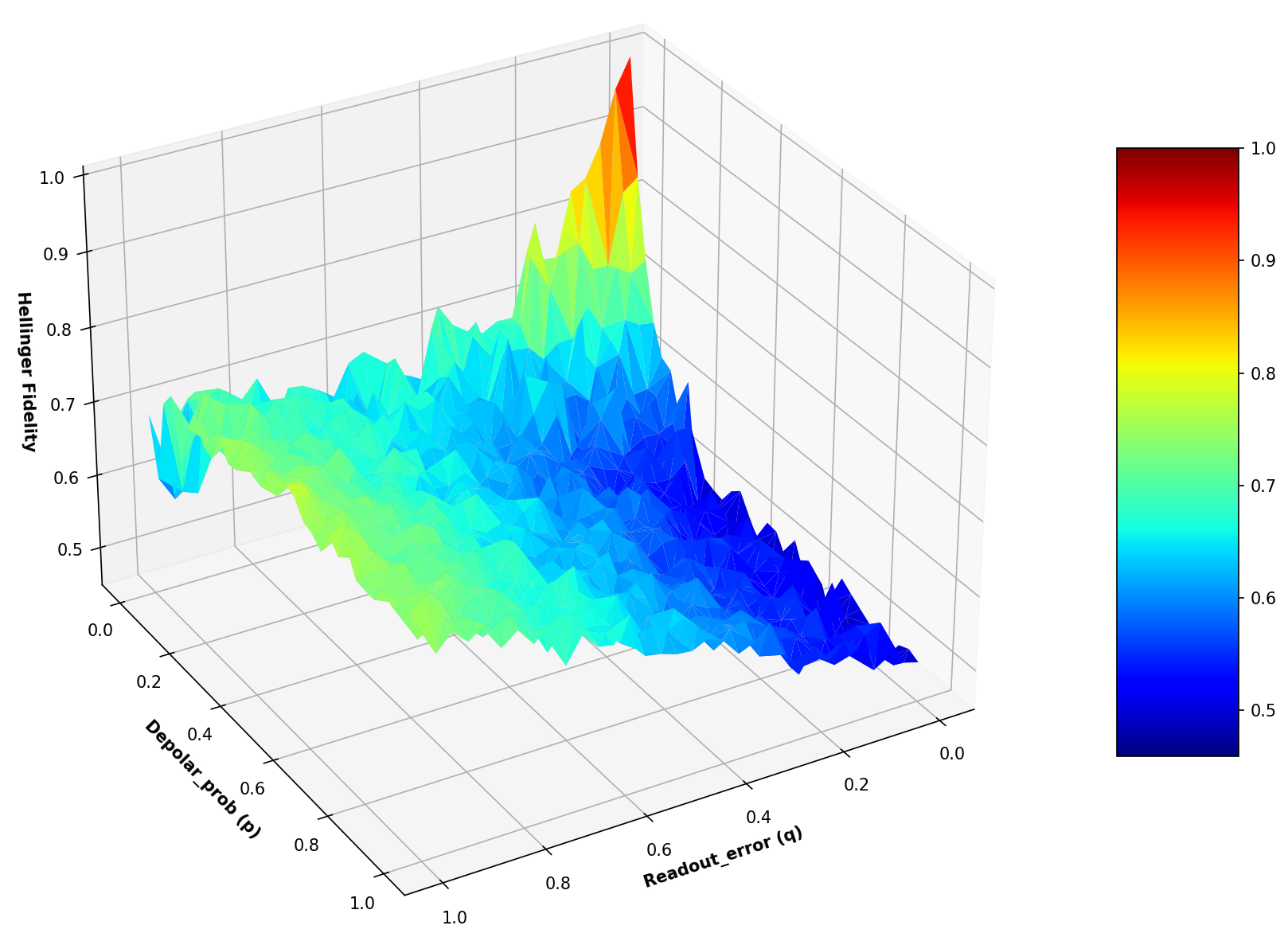}%
    }\\
\caption{\label{fig:Hfildelity}Example numerical simulations measuring the Hellinger fidelity, $F_H$, instead for each of the four schemes for three qubit circuit given a single random initial state. Generated surface is composed of 1024 simulation data points each run with 1024 shots.}
\label{num_surfaces_hellinger}
\end{figure}

 In Figure \ref{fig:Hfildelity}, we plot how $F_H$ varies with $q$ and $p$ for an example three qubit case, and interestingly, the surface shapes match closely the structure we saw in Figure \ref{num_surfaces_2}. Thus, we see that while our expectation value of zero measurement as the standard for `successful state transfer' may not be directly related to the exact quantum state fidelity itself, it is nevertheless a useful measurement as it can serve as an analogy to some notion of fidelity (namely, Hellinger fidelity).
\label{appendix:c}

\end{appendices}
\bibliography{references}%

\end{document}